\documentclass[nofootinbib,prd,supersciptaddress,floatfix,preprint]{revtex4-1}
\usepackage{amstext,amsmath,amssymb,amsfonts,bbm, amsthm}
\usepackage{graphicx}
\usepackage{caption}
\usepackage{bbold}
\usepackage[toc]{appendix}

\usepackage{graphicx,color}
\usepackage{subfig}
\usepackage{verbatim} 
\usepackage{mathrsfs}
\usepackage{hyperref}
\newcommand{\kets}[1]{\left\vert #1 \right\rangle}
\newcommand{\beq}{\begin{equation}}
\newcommand{\eeq}{\end{equation}}
\newcommand{\bqr}{\begin{eqnarray}}
\newcommand{\eqr}{\end{eqnarray}}

\begin{document}
\author{Daniel Bedingham }
\email{daniel.bedingham@philosophy.ox.ac.uk}
\affiliation{Faculty of Philosophy, University of Oxford, OX2 6GG, United Kingdom}

\author{Sujoy K. Modak}
\email{sujoy@post.kek.jp}
\affiliation{KEK Theory Center, High Energy Accelerator Research Organization (KEK),\\ Tsukuba, Ibaraki 305-0801, Japan}

\author{Daniel Sudarsky}
\email{sudarsky@nucleares.unam.mx}
\affiliation{Instituto de Ciencias Nucleares, Universidad Nacional Aut\'onoma de M\'exico, M\'exico D.F. 04510, M\'exico}

\title{ Relativistic  collapse  dynamics  and  black hole information loss}
\date{\today}

\begin{abstract}
We  study a proposal for the   resolution  of   the   black hole information puzzle   within   the context of     modified  versions of quantum theory   involving  spontaneous   reduction of the quantum state.  The theories   of this   kind,   which  were  developed  in order to address the   so  called  {\it measurement problem} in quantum theory  have,  in the past,  been framed in a non-relativistic  setting and   in that    form  they were  previously  applied   to  the black hole  information problem. Here, and for the first time,  we  show  in a   simple toy model,    a  treatment of the  problem   within a fully  relativistic  setting.    We also discuss the issues   that the present analysis  leaves    
 as  open problems  to  be   dealt with in future   refinements  of the present  approach.
\end{abstract}

\maketitle

\section{Introduction}
 
The discovery that, according to quantum field theory in its   general   curved  spacetime  version,   black holes   should radiate away  their energy \cite{hawk} ,  has  had  profound  impact on   both our   understanding of   the  interplay  between  gravitation and  quantum physics --teaching us  for instance that  the   laws  of   black hole mechanics  are in fact  the laws  of  thermodynamics  when applied to    situations  involving black holes--   and   in   contributing to   our  realization that    there   is  much  that  we   still need to understand \cite{hawk2} .   Regarding  the latter we   are  referring, of course,   to what  is commonly known  as the  ``black hole  information paradox''.
There   have been  many     attempts to address this  issue on the basis of proposed  theories incorporating   quantum treatments of  gravitation, and   it is fair to say that  none of those  seem to  offer  truly  satisfactory  resolution.   For a   review  see for instance \cite{Mathur1}.

 In fact,  there is  even  a   controversy  as to whether there is  or  there is   not, a paradox  or  some open issue  
 that needs  confronting.    In  \cite{Okon2},  this  question   has  been  discussed   and  clarified. The basic   
 issue  that   seems to underlie the various  postures in this  respect is   associated  to   the   view  that   one 
 takes  regarding  the nature of the  singularity that  is  generically found  deep in the black hole interior.       If  one  
 views  this   singularity as  a fundamental   boundary of spacetime,  there is  in fact no paradox  whatsoever,  
 as one  can   say  that  information  either   is  ``registered  on"  or   else   ``escapes    through" that boundary
 {\footnote{More appropriately    one should  think of   adding a boundary  arbitrarily close to the singularity and  
 use  that to   be part of the   Cauchy  surfaces    where  one   studies  the nature of the  quantum states at late 
 times.}}.  On the other hand,  if one  views,  as  do most researchers   working  in  the various  approaches to 
 quantum gravity,  the singularity as  something that  must be  ultimately  ``cured"   by  an appropriate quantum 
 theory of gravitation,   in   the sense  of  replacing  it by something amenable to  treatment   by such  theories  
 and  not  as  any  kind of   essential boundary (for instance the   proposal   within Loop  Quantum Gravity  
 discussed in \cite{bh-singularity}),  one    must  explain   how to  reconcile  the   unitarity of  quantum 
 mechanical  evolution (a feature  that among other  things requires reversibility and   thus the preservation of 
 information)  with the  thermal  nature   of  the Hawking radiation.  Without  any   reasonable  reconciliation of 
 the  divergent  conclusions  one  would be   entitled  to describe the situation as  a paradox. 
 
  In order to   explore the   most explicit version of   the problem   it is  customary  to    consider  a black hole  
  formed by the   gravitational collapse  of  a  large   lump of matter  (with  mass   of the order of,   say, a  few 
  solar  masses)  characterized  by a pure quantum mechanical  state.  The problem  one faces   then is to 
  reconcile  the purity of the initial state   with 
the  thermal  nature of  the Hawking  radiation.  The   issue  has been    studied    extensively  (see for instance  the  nice reviews  
 \cite{sGidd92, aStr95,  Benachenhou:1994af},   or the   works  \cite{ST, RST})  as it  is  considered  one of the major  challenges of  contemporary theoretical physics 
 
The  approaches   that   have been considered   in the search for  such  a reconciliation  seem  to  be  relatively   limited. Faced with the  
fundamental assumption (*): {\it  The  validity of quantum field theory  in curved space-times,  at least, in regions  
where  curvature is far  from the Planckian regime}, these approaches  essentially represent  
variations of the following ideas: 

 1)  Somehow,   during  the  ``late time" part of the  black hole  evaporation,    the Hawking   radiation  is  not 
 truly thermal,  and  is  in  fact highly  correlated with the early-time  radiation  
 (which must remain thermal  due to (*)),   so that the  full state of the radiation  field   is  pure.
 
 2) The  black hole  evaporation is not complete,  and   modifications  associated   with quantum gravity  lead to 
 the formation of a stable remnant.  Such a remnant  would have   to  be in   a highly  entangled  state  with the    
 emitted radiation so that the  full state of  the   (radiation field  +   remnant),  is pure.
 
 3) The Hawking radiation is thermal  all the way until the  eventual  evaporation of the   black hole 
 but   the   information   somehow  crosses the region  where the singularity   would  have  been  found,  and that  is  now  described  in terms of the   quantum gravity  theory.
 
Alternative  proposals  might involve   some   combinations of the three proposals  above.   However it seems clear 
that, at least one of them should be able to account for the fate of  most  of the  information. That is,  if  none of 
them  can account  for  anything more than a  very small fraction thereof,    then the  three  alternatives together will not be  able 
to account  for more than a   slightly  larger fraction of the  full information that needs to  be recovered,   if  the 
process  is to be  compatible  with   the unitarity of quantum mechanics.

 The   fact is that each of these  3 alternatives  have  serious  drawbacks.
 
  1)    This   idea,  which is  generally  framed  within the context of  the  so called black hole complementarity  proposals \cite{COMPL},  has  been the 
  subject of recent   detailed studies which show,  based on the so  called  monogamy 
  of quantum   entanglement,  that  one of   the consequences  of  such   entanglement (even 
  forgetting for the moment the question of how   would such  correlations   be   generated)  would 
  be the formation of  ``firewalls"  \cite{FireWalls} (or regions   of  divergent energy momentum of 
  the quantum field) around the black hole   horizon.
 
 2)  Here  the issue  is  that  one  would  be postulating  the  existence of   peculiar  kinds of objects, the remnants,   typically with a mass of   few times   the Planck mass,  which  must  have an  enormous   number of  internal states,  essentially  as many as those  of a large  star.  That is   
 because  the   full state of  the  (radiation field  +   remnant)  must be  pure  while  the  reduced  density matrix  characterizing the radiation field  is thermal,  and   has  an energy content  of  few solar masses.  
  
 3)  This   alternative seems to  be  favored  by  researchers  working in Loop Quantum Gravity, 
 and   has  been    considered in some detail  in  \cite{Ashtekar1}.
  Here, there are two issues  that need to be clarified. First, one  needs to  explain  precisely how the  information   crosses the   quantum gravity  region  that replaces the  classical singularity,  in  particular given that  in 
  the  LQG context,   that region  seems  to   be  characterized  by signature changes in the  metric
  \cite{Bojowald}. Secondly, there  seems  to  an  even   more problematic  aspect of this proposal:   The   fact that,  after the   complete   evaporation of  the  black hole, the  information missing  in the thermal  radiation  would   have to  be encoded in the  quantum gravity degrees of freedom (DOF), which however  would  have  an   essentially  vanishing  associated  energy. That  is,  the   quantum  gravity  DOF  would  have to  be entangled   with the Hawking  radiation in  such  a way that   the  complete state of  the quantum gravity  sector   plus the quantum matter field  sector,   would  be pure, and yet  the  energy   would  be   essentially  all  in the  radiated  quanta   of the field.

We  must  note that  there have been other proposals   such as those  considered in   \cite{wormhole, planck-star, Mathur2}   but  we  feel it  is  fair to  say that none of  these   has gained   any  kind of universal   acceptability   within the   community interested in the   issues, as  each  faces   some  difficulties  of its  own.

 We  want to  explore  a possible resolution of the paradox,  by  assuming that QG   would indeed  replace  the singularity by something   else,   suitably described in terms of  the fundamental DOF of   such a  theory,  but   that  quantum theory  would have to be modified  along the lines  of  the proposals put forward to  address the ``measurement problem"  by treating collapse of the wave function as a physical process,  occurring    spontaneously and independently of   ``observers  or measuring devices" , and  that the  corresponding   modification is  such,   as suggested  in \cite{Okon1, Okon2},  that  essentially  all   the  initial  information is  actually lost. 
      
      The first aspect  we  must   note about the  general proposal is  that its  setting is within the  general   context of semi-classical gravity. That is,  a scheme  where  the gravitational degrees of freedom  are treated using a   classical  spacetime metric,  while the matter degrees of freedom  are treated  using the  formalism of quantum field theory in  curved  space-times\cite{QFTinCS}.   The  first reaction of  many people   towards this  is to  cite the paper \cite{Page}  which  supposedly   rules out the viability of  semi-classical gravity. 
      Here  we must first   point out the  various  caveats  raised  about   such  a conclusion  in  \cite{Carlip}  and  note,   in particular,  that the  work in \cite{Page}  centered mainly in the    consideration of a  formulation in   which  quantum  theory   did not  involve  any sort of collapse of the quantum  state,  a  situation that contrasts  explicitly with  what we  will be   focusing on.   
 
  The  other  point noted in \cite{Page}    is  that   if  one   wants to consider  semi-classical gravity  together   with a version of  quantum  theory   involving  the  collapse of the quantum  state, one   faces the  problem  that the  semi-classical Einstein equation cannot   hold   during a collapse  simply because  Einstein's  tensor is  by construction  divergenceless while the expectation value of   the energy momentum tensor would  generically  have  a non-vanishing  divergence.

The point is that    one can  view  semi-classical gravity,  not as   a fundamental theory, but as  providing a 
suitably approximated   description  in limited  circumstances, something akin to say the hydrodynamical 
description of  a fluid   which,  as  we  know,  corresponds   only to the  description   of something that   
at a deeper level needs to be described in terms of molecules moving and  interacting    among 
themselves  in rather complex  ways.  Following the  analogy,  we  view  the metric  description of 
gravity and the characterization  of  the matter sector  using   quantum field theory  (and  connected  to   gravity  via  
Einstein's  semi-classical equations)   just as   an approximated  description of limited  validity.   
 In fact  this  is    a  point of view  that   has  been explored in the cosmological setting to  deal  with 
 certain difficulties  that arise in the  inflationary  cosmological   account for the  emergence of the 
 seeds of  cosmic  structure \cite{Collapse-and-Inflation}.  The introduction of dynamical collapse  
 within the general framework  can be treated    in a scheme   where  one  allows    an instantaneous 
 violation of the equations,  in association  with the collapse of the  quantum state  taking place on a  
 given  spatial hypersurface,  and   in analogy  with Israel's  matching conditions \cite{israel} requiring    continuity 
 of   the  metric  across  such a hypersurface. The  details of that formalism  were first described   in    
 \cite{Alberto}.  We   will not   discuss   these  issues further here as they  have been thoroughly 
 treated in   the above reference and   also  in the previous  works    by  some of  us  on the black hole 
 information  problem \cite{Sujoy 1,  Sujoy 2}.

    Here it is  worth  pausing  to  reconsider in more  detail   certain   aspects of the  
    discussion  around  the  issues  of  energy content.  The   setting  of the   discussion is   that of  
    black  holes  in asymptotically flat space-times.   For these space-times  we have a  well defined 
    notion of ADM  mass which   is taken as  the covariant  measure  of the energy  content  of the 
    spacetime,  and the quantity that is conserved   in the sense  that   the evaluation of the ADM  mass 
    gives the same  number  when computed  using any   Cauchy  hypersurface  $ \Sigma$ (which  in the 
    extended  spacetime ends  at $i^0$). 
    
 Moreover   as the spacetime  extensions also include    the   regions   ${\cal J}^+$  and   ${\cal J}^-$   
 i.e  asymptotic  future  and  past  null   infinity   respectively)  one  can use the notion of  Bondi mass   
 associated  with any  hypersurface $ \Sigma'$ ending   at  a section 
  $p \in {\cal J}^+$, (that  is,   $ \Sigma'$, together with  the  segment   of ${\cal J}^+$ starting at   
  $i^0$ and  ending  at  $p$  would   be  a   Cauchy  hypersurface).   The   point is that the Bondi mass  
  at $p$   should be  equal to the  initial ADM  mass of the spacetime   minus the amount of  energy  that 
  has  been radiated to  the segment  of ${\cal J}^+$ starting at   $i^0$ and  ending  at  $p$ .
 
 \begin{figure}
\centering
\includegraphics[scale=0.5]{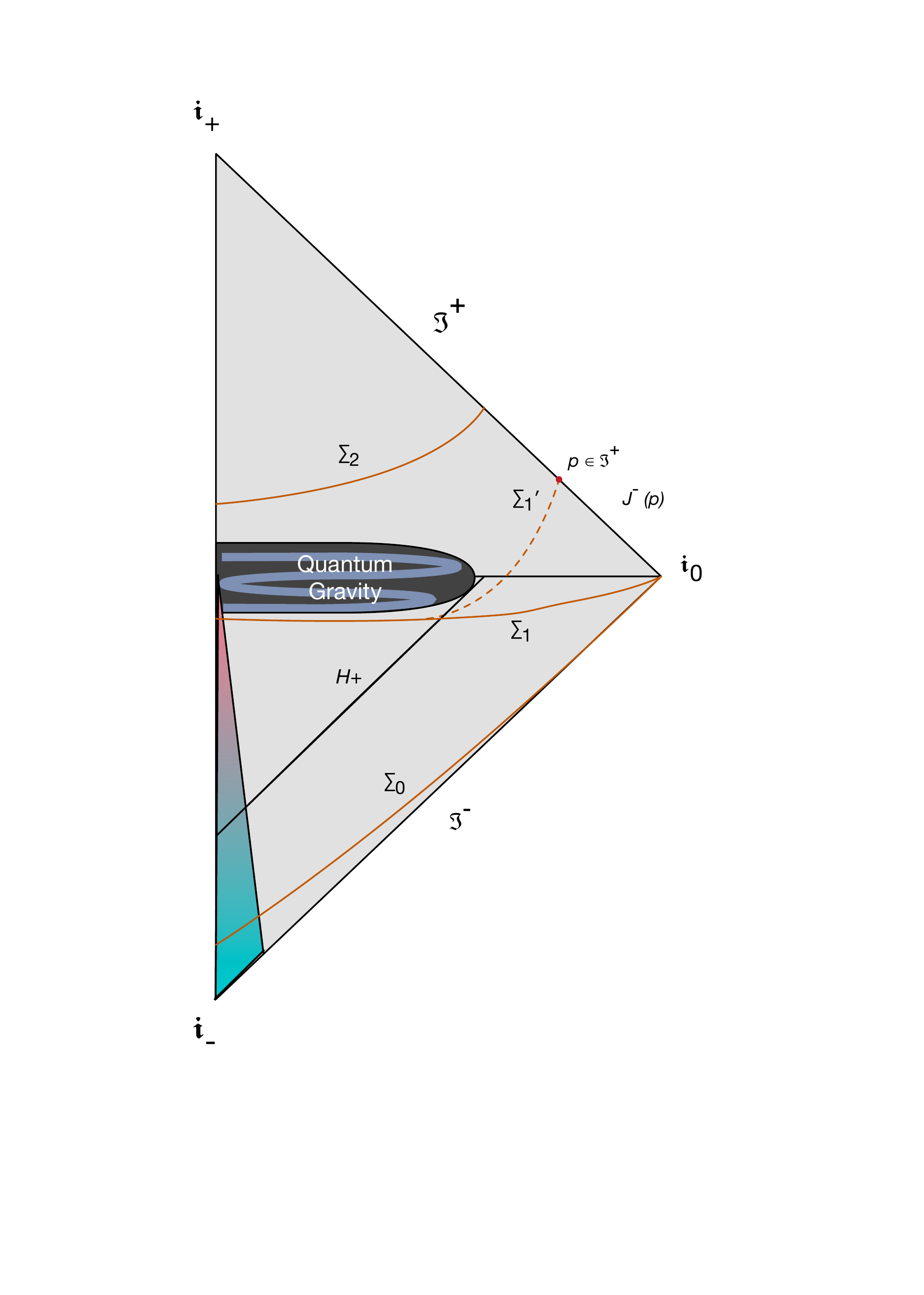}
\caption{Penrose diagram for the black hole spacetime } 
\label{PENROSE4D}
\end{figure}

  We  must now clarify  in  what sense  are we  going to  be using the  notions  of ADM 
   mass and Bondi  mass    as  being associated  with  Cauchy  hypersurfaces    $ \Sigma$ and  partial  Cauchy  
hypersurfaces   $ \Sigma'$.  Let us  concentrate for a moment  on the spacetime  that lies well to 
the past of the singularity (or the  would be  singularity that presumably is cured  by QG).    The point 
is that,   although   formally the expression for  say the ADM  mass  is   associated  with an  ``integral 
at infinity"  the  behavior of the metric  variables  at infinity is  conditioned  by the energy momentum  
content associated   with  the matter fields   by the   Einstein equation.   In other words   we can  
compute  the ADM mass using the   Cauchy data  for the  gravitational  sector  on  $ \Sigma$, data 
 which  are  tied to the   energy momentum  of the matter fields through the  Hamiltonian and 
Momentum constraints.  In  that regard  we  might  want to understand how the various   components  of the  matter field  contribute to these constraints in each one of the  hypersurfaces  in question.  
 We  can say, for instance, that  associated  with the initial set up, we have a Cauchy  hypersurface  $
  \Sigma_0$ (see Fig. \ref{PENROSE4D})  where   we have  a  large   lump of matter    with a spacetime that is only  very mildly 
  curved   and   characterized  by a pure quantum mechanical  state of the matter fields, the  ADM mass 
  is  $ M^{ADM}_{\Sigma_0}$ which, as  we  noted, is of the order of a few  $M_{Sun}$. In that case  
  we   would say that the  energy of the spacetime,  is represented   almost  completely by that 
  encoded  in  the energy momentum  on the   matter  fields.  At  relatively late  times,  but still at the 
  past of the  singularity,  we    might consider  a Cauchy  hypersurface $\Sigma_1$, that   starts at 
  $i^0$   stays  close to ${\cal J}^+$      and finally enters the horizon and   ends  at the center of the   
  gravitationally collapsed    lump of matter  which   at  this  stage  is  well  within the   black hole 
  horizon.    Alternatively,   we might  consider  deforming  $\Sigma_1$ into  a  hypersurface that ends  
  on the  section  $p\in {\cal J}^+$  which we   call  the  hypersurface    $\Sigma_1'$ ( this  
  hypersurface  is,  of course,  not a Cauchy hypersurface).
  
If we  want now, to   account for the  energy  content in terms of data  on $\Sigma_1$,  as   
represented by 
$ M^{ADM}_{\Sigma_1}$ (which should be equal to $ M^{ADM}_{\Sigma_0}$), we would have to say  
that  there   is   a very important  component  of the  energy   content,
corresponding to the  energy momentum tensor    of  the  outgoing   Hawking radiation,  located   
in the         part of   $\Sigma_1$   which  lies on the  region  exterior to the horizon, while  the 
energy  contained in  the    original   lump of matter  
has  been   red-shifted  by the  gravitational  potential associated  with the black hole,   and at the 
same time there is  a  negative contribution  to   the energy content   associated with  the  in-falling 
counterpart   of the   Hawking radiation, which might be  considered  as also  lying  in the proximity of 
the   intersection of the  event   Horizon with $\Sigma_1$.    The situation is   depicted,  for the realistic  4  dimensional case   in  figure \ref{PENROSE4D}, and  for the   2  dimensional CGHS model  in \ref{cghs}.

In terms of   $\Sigma_1'$   we  would  say  that     we  need to  account  for the relatively small value
of the  Bondi  mass at   its   endpoint $p\in {\cal J}^+$,  a value that  is   obtained  by  subtracting from 
$ M^{ADM}_{\Sigma_0}$  the  energy carried  
away  by the Hawking  radiation that has reached   ${\cal J}^+$  
to the  past of $p$.
That   small   value of the  Bondi mass  would, in turn, be  accounted for,  in terms of  the  
data on   $\Sigma_1'$,    as resulting  from the  red shifted  energy of  the   original   lump of matter,  
and  the   negative contribution  associated with  the  in-falling counterpart  of the   Hawking radiation.

We  note that  the   situation   on    $\Sigma_1'$, as  far as  energy is concerned   is  very similar to that 
on  any  hypersurface characterizing the situation  well after the evaporation of the black hole  such as $\Sigma_2$  in  the  accompanying  figure.

Now,  we might  in  a similar  way,  want  to    consider  the fate  of   the information  in the picture 
above.  That is  we want to consider  how  is   the  information,   that was present in the  quantum 
state  characterizing the   initial state of  system  at $\Sigma_0$,  accounted for,   in  terms of the   
quantum state   characterizing  the  system  on  $\Sigma_1$?   The point is that,   by   deforming  $
\Sigma_1$   into   $\Sigma_1'  \cup    J^-(p)$  (where $  J^-(p)$  is  the part 
of  ${\cal J}^+$  to the  past of $p$),     we  can describe  the state,  as  an entangled  state,  which on    
$ J^-(p)$ is  just the Hawking thermal  radiation,    and  on    $\Sigma_1' $  is  also a   highly  
mixed  density matrix,  but   such that  the  complete state is pure. 

The point that  we  want to make here is that   the  above situation   seems to  be afflicted    by the   
same troublesome  aspects  which  were raised  in the  context of   the alternative  3) above. That is, 
the  state  of the system   on  $\Sigma_1' $  is  one  with an  enormous   number of  degrees  of  
freedom and  yet  a  very small value of the energy.  It  seems  therefore that   if  we   have   an 
explanation for the loss of  information   in the black hole  evaporation that  relies  on  losses  
associated  only  with the  QG  region (i.e.  losses that,   in the  GR language,  would  be   described  as   
produced  by the singularity)  we  would  still face the uncomfortable  aspects  that  lead to the 
rejection of alternative  3)    above, but  this   time  associated with the situation   prior to the 
singularity (i.e. for instance   the situation on $\Sigma_1'$ ) .  We  think  that in  addressing the problem   via   the  
introduction of   modifications of quantum theory  that  last  problem  is   dramatically ameliorated.

  The  notion that  one  could  learn   to  live  with   information being  lost  in  association  with  the  
  evaporation of  black holes,  has  been    considered  in some detail  in \cite{Wald-Unruh}, where  the 
  earlier  arguments \cite{Peskin}  indicating that such proposals would  necessarily  involve large 
  violations of known conservation laws   or  dramatic   violations of causal behavior have  been 
  dispelled.  In that  analysis, however,  the  resulting picture   seems to  be that   all the information 
  loss  occurs  at  the singularity,  i.e at the region that would be  described in non-metric  terms in a  
  quantum theory of gravitation,  while,   in the regions  where the metric  description would  be  
  appropriate,  one would have exact quantum mechanical  unitary evolution. Our view  is that such an  
  approach  offers  a less unified  view of physics than the one  we   are advancing, and   that,  as  a  
  result,  it might be  more vulnerable to  questions of self consistency.  For  instance,  if we  accept that  
  there are  violations  to the quantum mechanical   unitary  evolution,  but that those only occur  in 
  connection  with black hole evaporation,  we   might   have  problems,   with an  ultimate   quantum 
  gravity theory,  that   can be expected to include the possibility of   processes  involving   virtual  
  black holes.   In other  words,   we  might have to face  up to the expected  result of such a theory 
  indicating that   all physical processes  must involve contributions from  all  possible intermediate  
  states according  to   a  path integral formulation   of the process, and that   those intermediate 
  states  would involve also virtual  black holes, which in turn  would have   associated  violations of  
  unitarity.  As  first  discussed in \cite{Okon1}, this    kind of 
  problem   seems  less likely to arise  in a 
  more  unified  version  we are considering,  where  the violation of unitarity  is 
    an integral   aspect of  
  the   fundamental physical  laws  as envisaged in the various proposals for  modifications of quantum 
  theory that have been made in the context of   the  search for a resolution of the  so called  
  ``measurement problem" \cite{Pearle:76} - \cite{moreCSL}.  
  
   It  is    worthwhile reminding the reader that  the  so  called  measurement problem in quantum  theory  is tied to the 
   interpretational difficulties that   arise  when one   does not  want to   introduce,  in the treatment, some artificial   classical/
   quantum  cut (sometimes  presented  as  a macro/ micro physics  cut)  and    instead, one  wants to consider that  everything,  
   including potential observers  and  measuring apparatuses, should  be treated  in a quantum mechanical  language.  We  direct 
   the reader to  the works   in \cite{measurement}  for a good  overview,  to \cite {More-measurement}  for  a   more  
   extensive collection of postures, or to \cite {Mau:95} for  a  very  clear   recent analysis.   The relevance of this issue to  the problem at hand  can be   seen  
from the fact that   quantum theory  calls  for  purely unitary  evolution  only when  one  is dealing with a  completely isolated  
system  in which all   degrees of freedom  are treated  quantum mechanically.   
 
  Many  proposals to deal  with the general interpretational difficulties of quantum theory, and  in 
  particular with the  measurement  problem   have  been  considered   since the  inception of  the theory 
  \cite{Interpretaciones} and  there is  also  a  good  body of literature devoted to the  problems  of many of these proposals \cite{critiques}.   
  
   We   want to focus  on the   dynamical reduction theories   which involve  a  modification   of 
  quantum  dynamics  involving  spontaneous reduction of the quantum state. These  kind   of proposals  
  are   commonly known as   ``Collapse   Theories"  and have   a  rather long  tradition. For  recent reviews see    \cite {collapse model 
  review}. Recently    various relativistic   versions of    spontaneous  dynamical collapse  theories  have
    been  put forward   \cite{BED1, BED2}, \cite{relativistic collapse models}.

 We  could not end this introduction  without  acknowledging the strong inspiration  that we have drawn  
 from  R. Penrose's   discussions connecting   foundational aspects of quantum theory to  ideas about the 
 nature of quantum gravity \cite{Penrose, penrose-new1}.  In fact  in a very  early  analysis \cite  {Penrose-BH-Collpase}   
 R. Penrose  noted that  if   one  wanted to  obtain   a self-consistent picture   of  a   situation involving 
 thermodynamical equilibrium     that  included  black holes  one  would need to  have a theory of quantum 
 mechanics that incorporated  some  violation of unitarity in  ordinary conditions  (involving no  black 
 holes). We view  that analysis as   providing   further    support   for  our  approach in  contrast to  those   
 where  violation of unitarity is  only associated  with  the  singularity  in evaporating  black holes.  Some of   Penrose's recent works \cite{penrose-new2} on such issues, 
 are also relevant, in a broader sense, to our  general views  underlying this proposal.
  
We  will  present  here  an  concrete version of   the above  approach   based on the   theory developed in  \cite{BED1, BED2}. The  article  is  organized  as follows:  In  section  II   we  present a brief description of the   CGHS   2  dimensional model   of  black hole  formation and evaporation, in  section III  we present  a  relativistic model of  dynamical collapse, section  IV  describes the   general  setting  in  which  we   will  put together  the  two   elements    previously  described,  and     in section  V   we   will  use  
them  to  describe the evolution of   the quantum state of  the mater field    thus accounting  for the  loss of information.  In  section VI   we   discuss  some  subtle points   regarding the  energetic   aspects of the   proposal  and   we  end    in  section  VII   with the general conclusions   indicating what has  been  achieved and   what  would  need be  left  as    issues   for further  research.   We  have  added      two  appendices  for the interested  reader  convenience:  Appendix  A    discusses  in detail the 
foliation independence of the  proposal,   exhibiting its   general covariance, and   Appendix  B   presents    in   some    detail  the  manner  in  which   the   delicate  issue regarding the  expectation of   unbounded   energy   creation   is  resolved  by the introduction of the {\it pointer field}.


\section{ Review of the CGHS  Model}
\label{CGHSModel}

The  two dimensional   model, first introduced  by  Callan-Giddings-Harvey-Strominger (CGHS)  \cite{CGHS92} involving  black hole formation  is  a very convenient   toy  model  for the   study of  issues  related to the  formation and evaporation of  two dimensional black holes. 
  
\begin{figure}
\centering
\includegraphics[scale=0.5]{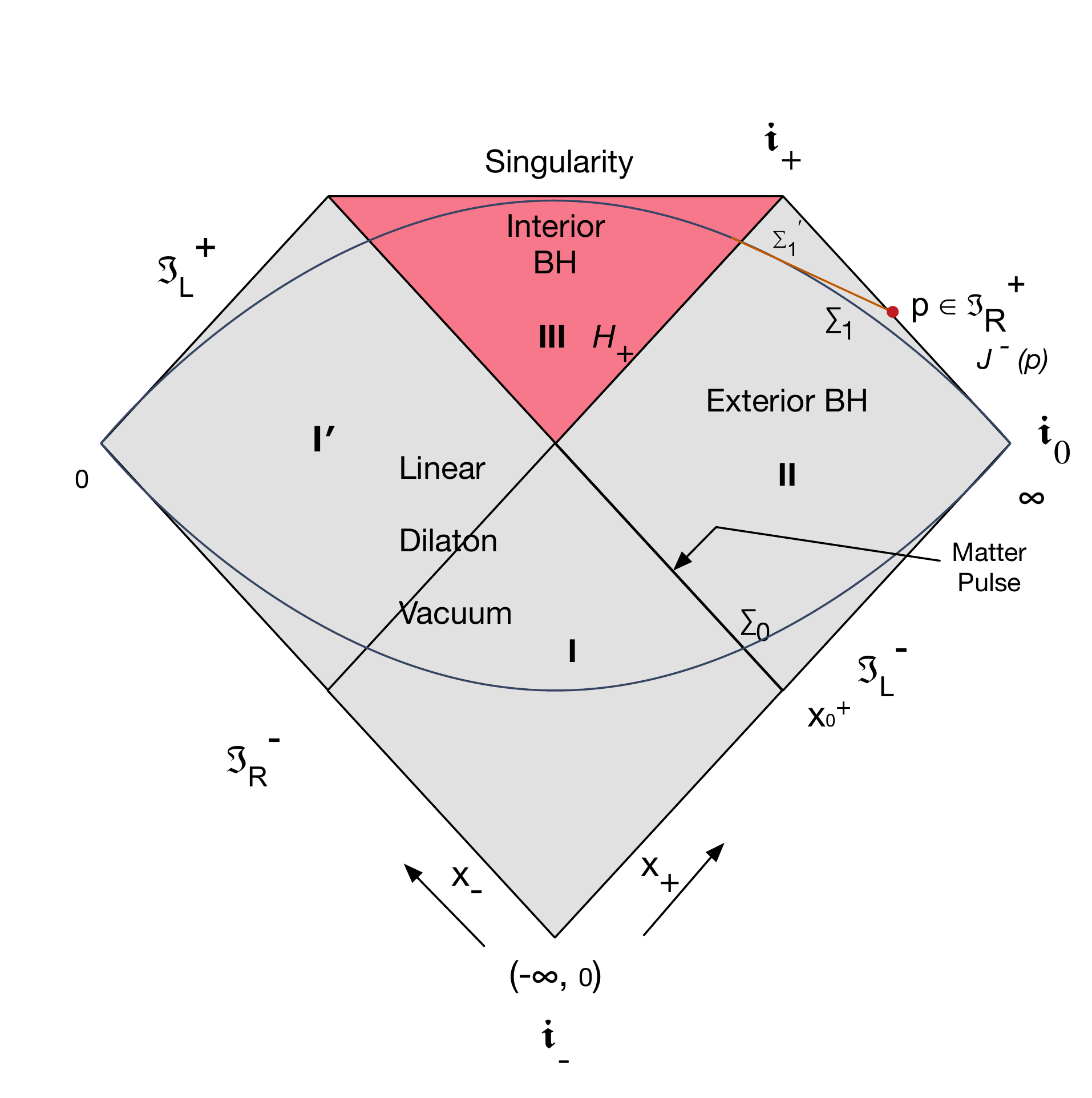}
\caption{Penrose diagram for CGHS spacetime.}
\label{cg}
\end{figure}

We now review   the basic  features  of this  model. For more details we refer the reader to \cite{aFabjNs05}. The CGHS action is
\begin{align}
{ S=\frac{1}{2\pi}\int d^2x\sqrt{-g}\left[e^{-2\phi}\left[R+4(\nabla \phi)^2+4\Lambda^2\right]-\frac{1}{2}(\nabla f)^2\right]}, \nonumber 
\end{align}
where  $R$  is the Ricci scalar for the  metric $ g_{ab}$, {$\phi$} is the dilaton field, considered in this model  as   part of the gravity sector, { $\Lambda^2$} is  a cosmological constant and { $f$} is a  scalar  field,  representing  matter. The  solution corresponding to the CGHS model is shown in Fig. \ref{cg}.
 It corresponds to a  null shell of    matter  collapsing   gravitationally   along the world line   $ x^+ =x^+_0 $   and leading to the formation of  a black hole. For  $x^+ < x^+_{0}$,  this solution   is  known as the dilaton vacuum (region I and I'). The metric is found to be
\begin{align}
{ds^2=-\frac{dx^{+}dx^{-}}{-\Lambda^2 x^{+}x^{-}},}
\end{align}
 which is flat,  whereas for    $x^+ > x^+_0$ the solution     is described 
by the black hole metric (region II, III)  represented by
\begin{align}
ds^2=-\frac{dx^{+}dx^{-}}{\frac{M}{\Lambda}-\Lambda^2 x^{+}
(x^{-}+\Delta)},
\end{align}
   where  $\Delta = M/ \Lambda^3 x_0^{+} $. Here  (null) Kruskal-type  coordinates ($x^+, x^-$) are useful to describe the global structure of the spacetime. On the other hand, for physical studies involving Quantum Field Theory (QFT) in curved 
spacetime, it is convenient  to use  special  coordinates for the   various regions. 
In 
the dilation vacuum  region,  the natural coordinates are $y^+ \equiv
\frac{1}{\Lambda} \ln(\Lambda x^+) , y^- \equiv  \frac{1}{\Lambda} \ln(-{\frac{x^-}{\Delta}})$, and thus the  metric  can   be  expressed as $ds^2=-dy^{+}dy^{-}$ with $-\infty < y^{-} <\infty;~ -\infty <y^{+} < \frac{1}{\Lambda}
\ln(\Lambda x_0^+)$. 

In  the BH  exterior (region II),  a    natural   set of 
coordinates   is   provided  by $\sigma^+ \equiv \frac{1}{\Lambda} \ln(\Lambda x^+) 
, \sigma^-  \equiv -\frac{1}{\Lambda} 
\ln(-\Lambda(x^- + \Delta))$, so that   the metric in this region is $ds^2=-\frac{d\sigma^{+}d\sigma^{-}}{1+(M/
\Lambda)e^{\Lambda(\sigma^{-}-\sigma^{+)}}}$ with $-\infty < \sigma^{-} < \infty$ and $\sigma^{+} > 
\sigma_{0}^{+}= \frac{1}{\Lambda} \ln(\Lambda x_0^+).$  
In order to  exhibit  the asymptotic flatness, we  express 
the BH metric in Schwarzschild-like coordinates {($t,r$)} which are defined through the  implicit  formulas   {$\sigma^{\pm} =t \pm \frac{1}
{2\Lambda} \ln(e^{2\Lambda r} - M/\Lambda)$} so that we get ${ ds^2= -(1-\frac{M}{\Lambda}
e^{-2\lambda r}) dt^2 + \frac{dr^2}{(1-\frac{M}{\Lambda}e^{-2\lambda r})}}$. The   temporal and  spatial Kruskal coordinates 
$T =  (1/2) (x^+ + x^- + \Delta),~X = (1/2)(x^+ - x^- -\Delta)$ can be related to Schwarzschild-like time $t$ and 
space $r$ coordinates,  through  $\tanh(\Lambda t) = T/X$ and $-\frac{1}{\Lambda^2}(e^{2\Lambda r} - M/
\Lambda) = T^2-X^2$.

  Now we consider  the  quantum treatment of the matter field  $f$.  We   will  consider    
  the  null  past   asymptotic   regions { ${\cal{J}}^{-}_{L}$} and {${\cal{J}}^{-}_{R}$}  as the {\it in} region  and the black hole (exterior and interior) region as the asymptotic {\it out} region. 
 
  In the {\it in} region,  the field operator can be expanded as $\hat{f}(x) = \sum_{\omega}(\hat{f}_{\omega}^{R}(x) + 
  \hat{f}_{\omega}^{L}(x)),$ where $\hat{f}_{\omega}^{R/L} = \hat{a}_{\omega}^{R/L} u_{\omega}^{R/L} + 
  \hat{a}^{R/L \dag}_{\omega} u_{\omega}^{R*/L*}$,
and the basis of functions (modes) are:
$
{u_{\omega}^{R}=\frac{1}{\sqrt{2\omega}}e^{-i\omega y^{-}}}
$
and 
$
{u_{\omega}^{L}=\frac{1}{\sqrt{2\omega}}e^{-i\omega y^{+}}},
$
with {$\omega>0$}. The superscripts {$R$} and {$L$} refer to the right and left moving modes respectively. 
These modes define   the   bases of  field  quantization  and thus  the  right {\it in} vacuum ({$\kets{0_{in}}_{R}$}) and  the left {\it in} vacuum  ({$\kets{0_{in}}_{L}$}) 
whose tensor product ({$\kets{0_{in}}_{R}\otimes \kets{0_{in}}_{L}$}) defines our {\it in} vacuum. 

  As  is well known,  one might  also proceed to the  construction of the   field  theory in   terms of modes that  are natural   in  the {\it out} region by   expanding  the field operator  $ \hat f$ in terms of the complete set of modes having support both outside (exterior) and inside (interior) the event horizon. Once more  we  can   write  the field operator in the form $\hat{f}(x) = \hat{f}^{R}(x) + \hat{f}^{L}(x)$  where  
\begin{align}
 \nonumber
\hat{f}^{R/L}(x) = \sum_{\omega}\hat{b}_{\omega}^{R/L} v_{\omega}^{R/L} + \hat{b}^{R/L \dag}_{\omega} v_{\omega}^{R*/L*} 
 + \sum_{\tilde\omega}\hat{b}_{\tilde{\omega}}^{R/L} v_{\tilde{\omega}}^{R/L} + \hat{b}^{R/L \dag}_{\tilde{\omega}} v_{\tilde{\omega}}^{R*/L*}. 
\end{align}
In the above  we have used the convention whereby modes and operators with and without tildes correspond to the     regions  inside 
and outside the horizon, respectively. 

    For the  mode  functions  in the exterior to the  horizon  we use: 
${v_{\omega}^{R}=\frac{1}{\sqrt{2\omega}}e^{-i\omega\sigma^{-}}\Theta(-(x^{-} + \Delta))}$
and
$ {v_{\omega}^{L}=\frac{1}{\sqrt{2\omega}}e^{-i\omega\sigma^{+}}\Theta(x^{+} - x_0^{+}).}$
Similarly  we  can choose the  set of modes  in the black hole interior,    ensuring  that the basis of modes in the {\it out} region is 
complete. The left moving modes are kept the same  as before  (since these modes  travel from  the  black hole  exterior 
to the interior), while  for  the right moving mode we take:
$
{\hat{v}_{\tilde{\omega}}^{R}=\frac{1}{\sqrt{2\tilde{\omega}}}e^{i\tilde{\omega}\sigma_{in}^{-}}
\Theta(x^{-} + \Delta)}
$.
  Following \cite{cghs-more}, we now replace the above delocalized plane wave modes by a complete orthonormal set 
  of discrete wave packets modes, given by $v_{nj}^{L/R}=\frac{1}{\sqrt{\epsilon}}\int_{j\epsilon}^{(j+1)\epsilon}d\omega e^{2\pi i\omega n/\epsilon}
v_{\omega}^{L/R},$ where the integers $j\ge 0$ and  $-\infty <n<\infty$. These wave packets are naturally peaked about $\sigma^{+/-}= 2\pi n/\epsilon$ with width $2\pi/\epsilon$  respectively. 

 The   next  step in our  analysis  is  to consider  the  Bogolubov transformations.  In our
case, the relevant non-trivial one  refers to  the right moving sector, and  the  corresponding  transformation 
 from {\it in} to {\it exterior} modes is  what  accounts  for the Hawking radiation. We note that  the initial   
 state,   corresponding  to the   vacuum   for the right moving modes and the    left moving pulse    
  forming  the black hole    $|\Psi_{in}\rangle =|0_{in}\rangle_{R} \otimes |Pulse\rangle_{L}$   
  can be  written as: 
\begin{equation}
{\cal N} \displaystyle\sum_{F_{nj}} C_{F_{nj}} |F_{nj}\rangle^{ext} \otimes |F_{nj}\rangle^{int}\otimes |Pulse\rangle_{L}, \label{inst}
\end{equation}
where  particle states { $F_{nj}$} consist of arbitrary but {\it finite} number of particles, {$\cal N$} is a 
normalization constant, and the coefficients {$C_{F_{nj}}$}'s are determined using the Bogolubov 
transformations. Their  explicit expressions can be  seen in \cite{aFabjNs05} .

 It is well known that, if one  ignores the  degrees of freedom of the  quantum field lying in the   black 
 hole  interior, and describes   just  the  exterior   DOF of  freedom,  one     ends up (partially)   describing the 
 state  in terms   of  a density matrix. That is,   one  obtains  the  reduced  density matrix   by    tracing   over 
 the interior  degrees of freedom (DOF),   and  in this  case  one   ends    up,    with  a  density  matrix  corresponding to a thermal  
 state.   Note, at this  point   this density matrix   represents, 
    in the language of \cite{dEspagnat} an {\it improper} mixture, as it arises after ignoring part of the system  which as  a whole  is in a  pure state.  We  will therefore  say that  
 what we obtain   at this point is an {\it improper thermal state}.  See   discussion in section \cite{Sujoy 2}  for a   more  exhaustive   discussion and analysis   of this  issue.  
 
  It is also  discussed in previous  works \cite{Sujoy 1, Sujoy 2}  that the task of accounting  for the information  loss 
  in  black hole  evaporation   within the    approach we   are   considering   requires  among other things showing   how,   as  the result of the   dynamics,   one  ends  up 
   with a {\it proper thermal state } (i.e.,  one that describes  an actual  mixed  state  that is  not  a partial description of  a pure  state)  starting with  an {\it initial  pure state}.


\section{Relativistic collapse: general formalism}
\label{CovColl}

For the purpose of presenting relativistic collapse models in generality we employ the interaction picture \cite{BF, BPP, relativistic collapse  ideas 2} in which the quantum state of matter $|\Psi_{\Sigma}\rangle$ is assigned to a space-like hypersurface $\Sigma$. As we advance the hypersurface $\Sigma$ to the future via some arbitrary foliation of spacetime, the state changes according to 
\begin{align}
i\frac{\delta|\Psi_{\Sigma}\rangle}{\delta\Sigma(x)} = \hat{\cal H}_{\rm int}(x)|\Psi_{\Sigma}\rangle,
\label{TOM0}
\end{align}
where ${\cal H}_{\rm int}$ is the interaction Hamiltonian density.  The functional derivative is defined as
\begin{align}
\frac{\delta |\Psi_{\Sigma}\rangle}{\delta\Sigma(x)} = \lim_{\Sigma'\rightarrow \Sigma}\frac{|\Psi_{\Sigma'}\rangle-|\Psi_{\Sigma}\rangle}{\Delta V},
\end{align}
where $\Delta V$ is the invariant spacetime volume enclosed by $\Sigma$ and $\Sigma'$  with $\Sigma\prec \Sigma'$ (meaning that no point in $\Sigma$ is to the future of $\Sigma'$). Covariance requires that $[\hat{\cal H}_{\rm int}(x),\hat{\cal H}_{\rm int}(y)] = 0$ for spacelike separated $x$ and $y$. This guarantees that advancing of the hypersurface across the points $x$ and $y$ is independent of the order in which this is done, and more generally it guarantees foliation independence of the state development.

The solution to (\ref{TOM0}) can be written as
\begin{align}
|\Psi_{\Sigma'}\rangle  = \hat{U}[\Sigma',\Sigma]|\Psi_{\Sigma}\rangle.
\end{align}
where $\hat{U}$ satisfies 
\begin{align}
i\frac{\delta \hat{U}[\Sigma,\Sigma_0]}{\delta\Sigma(x)} = \hat{\cal H}_{\rm int}(x)\hat{U}[\Sigma,\Sigma_0],
\end{align}
with initial condition $\hat{U}[\Sigma,\Sigma] = 1$. This can be formally solved to give
\begin{align}
\hat{U}[\Sigma_2,\Sigma_1] = T\exp\left[ -i \int_{\Sigma_1}^{\Sigma_2} \hat{\cal H}_{\rm int}(x) dV\right].
\end{align}
where $T$ is the time ordering operator, and $\Sigma_1\prec \Sigma_2$.

From time to time we suppose that the state undergoes a discrete collapse event associated to a spacetime point $x$. When the hypersurface $\Sigma$ crosses the point $x$, the state ceases for an instant to satisfy equation (\ref{TOM0}) and instead changes according to the rule
\begin{align}
|\Psi_{\Sigma}\rangle \rightarrow |\Psi_{\Sigma^+}\rangle  = \hat{L}_x(Z_x) |\Psi_{\Sigma}\rangle,
\end{align}
where $\hat{L}_x$ is the collapse operator at $x$ and $Z_x$ is a random variable which corresponds to the collapse outcome.{ One  normally  assumes that there  is  a fixed probability of a collapse event  occurring in any incremental spacetime region of invariant volume}. This results in collapse events which have a Poisson distribution  with  density   $\mu$,  in any unit volume of spacetime\footnote{As  we  will  see, in this  work  we  will assume that this quantity  can  depend  on  the local  spacetime  curvature.}.  This distribution of collapse events in spacetime is covariantly defined and makes no reference to any preferred foliation.

The collapse operators must satisfy the completeness condition 
\begin{align}
\int dZ |\hat{L}(Z)|^2 = 1.
\label{COMP}
\end{align}
This allows us to define the probability density for the outcome  $Z_x$, for a collapse event on the state $|\Psi_{\Sigma}\rangle$ at point $x$, by 
\begin{align}
\mathbb{P}\left(Z_x\right||\Psi_{\Sigma}\rangle) =
\frac{\langle \Psi_{\Sigma}||\hat{L}_x(Z_x)|^2|\Psi_{\Sigma}\rangle}{\langle \Psi_{\Sigma}|\Psi_{\Sigma}\rangle}
 = \frac{\langle \Psi_{\Sigma^+} |\Psi_{\Sigma^+}\rangle }{\langle \Psi_{\Sigma}|\Psi_{\Sigma}\rangle}.
\label{PROB}
\end{align}
The completeness condition ensures that (\ref{PROB}) is normalized. This formula corresponds to the 
standard formula for the quantum probability of a generalized measurement with measurement 
operator $\hat{L}_x$. The collapse outcomes thus occur with standard quantum probability.

In Appendix \ref{Ax:folindep} we demonstrate that if the following microcausality conditions hold, 
\begin{align}
[\hat{L}_x(Z_x),\hat{L}_y(Z_y)]= 0,
\label{COM2}
\end{align}
and
\begin{align}
[\hat{L}_x(Z_x),\hat{\cal H}_{\rm int}(y)]= 0,
\label{COM3}
\end{align}
for spacelike separated $x$ and $y$, then  (i) given a Poisson distributed set of collapse locations $\{x_j|\Sigma_f \succ {x_j}\succ \Sigma_i\}$ (with labels $j= 1,\ldots,n$, which give an arbitrary total ordering which respects the causal ordering of the spacetime) occurring between hypersurfaces $\Sigma_i$ and $\Sigma_f$, and a compete set of collapse outcomes at these locations $\{Z_{x_j}|\Sigma_f \succ {x_j}\succ \Sigma_i\}$, the state dynamics leads to an unambiguous and foliation-independent change of state between $\Sigma_i$ and $\Sigma_f$; and (ii) the probability rule specifies the joint probability of complete sets of collapse outcomes $\{Z_{x_j}|\Sigma_f \succ x_j\succ \Sigma_i\}$ independently of spacetime foliation, given only the state on the initial surface $\Sigma_i$.

The joint probability density  for the set  of   outcomes $\{Z_{x_j}|\Sigma_f \succ {x_j}\succ \Sigma_i\}$ can be determined from (\ref{PROB}) by 
 repeatedly making use of the definition of conditional probability, and is given by 
\begin{align}
\mathbb{P} \left(\left\{Z_{x_j}| \Sigma_f \succ x_j \succ \Sigma_i \right\}||\Psi_{\Sigma_i}\rangle\right)=
\frac{\langle \Psi_{\Sigma_f} |\Psi_{\Sigma_f}\rangle }{\langle \Psi_{\Sigma_i}|\Psi_{\Sigma_i}\rangle}.
\label{PROBn}
\end{align}
where $|\Psi_{\Sigma_f}\rangle$ depends on $\left\{Z_{x_j}| \Sigma_f \succ x_j \succ \Sigma_i \right\}$ as
\begin{align}
|\Psi_{\Sigma_f}\rangle = \hat{U}[\Sigma_f,\Sigma_n]\hat{L}_{x_n}(Z_{x_n})\cdots \hat{L}_{x_1}(Z_{x_1})\hat{U}[\Sigma_1,\Sigma_i]|\Psi_{\Sigma_i}\rangle,
\label{EVOn}
\end{align}
and where the choice of foliation $\Sigma_i\prec\Sigma_1\prec\cdots \prec \Sigma_n\prec\Sigma_f$ is arbitrary and corresponds to the arbitrary total ordering of $\{x_j\}$. At this point one can take the view that the resulting state histories with respect to different foliations are merely different descriptions of the same events \cite{relativistic collapse  ideas 1}. Alternatively, one can regard the collapse outcomes as the primitives of the theory from which the quantum state histories are derived.
 
The covariant form of the collapse dynamics together with the absence of any foliation dependence result in an adequate framework for a relativistic collapse model. To realize such a model we must propose a form for a collapse operator $L_x$ which satisfies the above requirements. We begin by choosing
\begin{align}
\hat{L}_x(Z_x) = \frac{1}{(2\pi\zeta^2)^{1/4}}\exp\left\{-\frac{(\hat{B}(x) - Z_x)^2}{4\zeta^2}\right\},
\label{L}
\end{align}
where $\hat{B}(x)$ is an, as yet unspecified  hermitian operator, and $\zeta$ is a new fundamental parameter. This collapse operator describes a quasi projection  of the state  of the system  onto  an approximate  eigenstate of $\hat{B}(x)$ about the point $Z_x$ meaning that, if the state  previous  to the  collapse   event  was represented in terms of eigenstates of $\hat{B}(x)$, the collapse effect is to diminish the relative   amplitude of eigenstates whose eigenvalues are far from $Z_x$ with respect to those that have eigenvalues close to  $Z_x$. The effect of many such collapses is to drive the state towards a $\hat{B}(x)$-eigenstate.

This collapse operator automatically satisfies the completeness condition. The microcausality conditions are satisfied if
\begin{align}
\left[\hat{B}(x), \hat{B}(y)\right] =0 \quad\text{and}\quad \left[\hat{B}(x), \hat{\cal H}_{\rm int}(y)\right] =0,
\end{align}
for space-like $x$ and $y$. We therefore propose  that, for  a theory of a  scalar field   such as the one  we  are considering  in this   work, 
\begin{align}
\hat{B}(x) = |\hat{f}(x)|^2,
\label{propose}
\end{align}
  where $\hat{f}(x)$ is the  scalar field operator.  This meets the above conditions for any interaction Hamiltonian given as a function of $\hat{f}(x)$. However, with this choice we face an immediate problem. If we calculate the average energy change in the field as a result of a collapse event we find
\begin{align}
\Delta E = \int dz\frac{\langle\Psi_{\Sigma}|\hat{L}_x(z)[\hat{H},\hat{L}_x(z)]|\Psi_{\Sigma}\rangle}{\langle\Psi_{\Sigma}|\Psi_{\Sigma}\rangle} = \frac{1}{2\zeta^2}\delta^{d-1}(0)\langle|\hat{f}(x)|^2\rangle,
\end{align}
for a $d$ dimensional spacetime where $\hat{H}$ is the Hamiltonian operator for the scalar field, and the final expression is the first order term in the large $\zeta$ expansion. This expression is infinite for a continuum spacetime. This could be ameliorated by a spacetime with fundamental discreteness (which  should not    enter in  conflict with  special  relativity \cite{CAUSALSET}). With discreteness length scale $l$ we could approximate $\delta^3(0)\sim l^{-3}$, and might then, by appropriate choices for the parameters of theory,  be able to construct a model in which the collapse of massive objects is sufficiently rapid whilst the average energy increase is sufficiently small to satisfy experimental lower bounds \cite{GERMANIUM}. (There are three parameters in this model: $\zeta$; the discreteness length scale $l$; and the spacetime density of collapse events $\mu$, which could   possibly be taken to correspond to the effective density of spacetime points, reducing the number of parameters to two.) Alternatively we propose the use of a new field to mediate the collapse process with the effect of preventing infinite energy increase. This construction is outlined in Appendix \ref{AX:AUX} where the effective collapse process satisfied by the scalar field is derived. In either the discrete space model or the auxiliary field model, the end result is a collapse model which drives the scalar field towards eigenstates of the operator $|\hat{f}(x)|^2$.  As in  any event  this is the  end result, we   will  be making  free use of it throughout  this paper. Thus  from here on  we will  mostly ignore the details of precisely how we deal with the problem of energy increase.

To understand the collapse rate we introduce the density matrix representation
\begin{align}
\hat\rho_{\Sigma} = \frac{|\Psi_{\Sigma}\rangle\langle \Psi_{\Sigma}|}{\langle \Psi_{\Sigma}| \Psi_{\Sigma}\rangle}.
\end{align}
A collapse event at point $x$ on the surface $\Sigma$ converts  the pure state into another pure state   with a  smaller uncertainty in $\hat B(x)$.  However  as the   specific  state  is    stochastically determined, it is  convenient to pass    to a description in terms of  ensembles. That  is  we consider  the   statistical mixture  representing the  ensemble of  a large number of identical  systems    characterized  by   the  same state   just before the collapse  event,   and    their  collective  change,   just   after    such  an event. This  is   thus  described  by:   
\begin{align} 
\hat\rho_{\Sigma}\rightarrow  \hat\rho_{\Sigma^+} = \int dz \mathbb{P}\left(z\right||\Psi_{\Sigma}\rangle)
\frac{\hat{L}_x(z)\hat\rho_{\Sigma}\hat{L}_x(z)}{{\rm Tr}[\hat{L}_x(z)\hat\rho_{\Sigma}\hat{L}_x(z) ]}
=\int dz  \hat{L}_x(z)\hat\rho_{\Sigma}\hat{L}_x(z).
\end{align}
This equation describes how the pure state at any stage is transformed into an ensemble of possible resultant  states, each element of which results from a particular value of the as yet unknown collapse outcome. The change in the statistical   density matrix  operator characterizing  the ensemble is  then,  
\begin{align}
\Delta \hat\rho_{\Sigma} = \hat\rho_{\Sigma^+} - \hat\rho_{\Sigma} = -\frac{1}{8\zeta^2}\left[|\hat{f}(x)|^2,[|\hat{f}(x)|^2,\hat\rho_{\Sigma}]\right],
\end{align}
in the large $\zeta$ limit. If we choose a foliation parametrized by $t$, with lapse function $N$ and spatial metric on the timeslices $h_{ij}$, and assume that there is a spacetime collapse density of $\mu$ then we can write
\begin{align}
\frac{d}{dt}\hat\rho_t = -i\int d^{d-1} x N\sqrt{h}[\hat{\cal H}_{\rm int}(x),\hat\rho_t]-\int d^{d-1} x N\sqrt{h} \frac{\mu}{8\zeta^2} \left[|\hat{f}(x)|^2,[|\hat{f}(x)|^2,\hat\rho_{t}]\right],
\label{PHICOLL}
\end{align}
where   $h$  stands for the determinant of the  components  of the metric  $h_{ij}$  in the coordinates $\lbrace   t, x_i \rbrace $. The  first term corresponds to  the unitary dynamics of the interaction Hamiltonian,  which  would  vanish  in the case of a  free  field  theory such as the one   we   are considering.

It  is  convenient at this point to consider  the evolution in  terms of  a basis   of    instantaneous  field  eigenstates   for the hypersurface  $\Sigma_t $ (corresponding to  a  leaf  of the foliation ,   $t= $ {\it constant}.    That   is  $|f\rangle_t$ are field eigenstates  on the  hypesurface $\Sigma_t $ (i.e.~states which satisfy $\hat{f}(x)|f'\rangle ={f'}(x)|f'\rangle $ ,  $ \forall  x  \in \Sigma_t$).  Such states form a complete basis of states    for  each  value of  $t$. 
  
   It  thus  follows that, 
\begin{align}
\frac{d}{dt}\langle f |\hat\rho_t|f'\rangle = -\Gamma[f,f']\langle f |\hat\rho_t|f'\rangle,
\end{align}
\begin{align}
\Gamma[f,f'] = \int d^{d-1} x N\sqrt{h} \frac{\mu}{8\zeta^2} \left[|f(x)|^2-|f'(x)|^2\right]^2.
\label{rate}
\end{align}
The coupling parameter $\gamma = \mu/8\zeta^2$ is usually 
 taken as a constant  but as  first  suggested in \cite{Okon1}  we  will   assume 
 it is  a   local function of   
 curvature scalars.   For  concreteness  we take   
$ \gamma  = \gamma( W^2)$    where $\gamma (.)$ is an  
increasing  function of its  argument,   $W^2 = W_{abcd} W^{abcd}$,  and $ W_{abcd} $ is
the Weyl  tensor for the   
spacetime  metric $g_{ab}$.   This  feature  ensures not only that the collapse  effects will be  much 
larger  in regions of    high  curvature   than in regions  where  the spacetime is   close  to flat  but it 
might  also be used to ensure that the    
completely flat regions   where  among other things  the matter content   corresponds to  the  
vacuum, the effects of 
collapse  disappear completely. In the two dimensional setting of the CGHS model, the Weyl curvature 
is zero and so as a 
 substitute we take $\gamma$ to be an increasing function of the scalar curvature $R$.

The upshot is that the particular relativistic collapse model determined by the proposal (\ref{propose}) leads to collapse in 
the ${\hat f}$ state basis at a rate given by (\ref{rate}).  The collapse process will not,  however, lead to a precise field  eigen-states, simply because the collapse   is  only  assumed to  narrow the   uncertainty,  and  the free dynamics of the field, will cause dispersion of the field state in competition with the collapse. In  fact,  what the  result  of  eq. (\ref{rate}) shows,  is  that  that states with different $ |f|^2$ are distinguished, rather than  states with different  $f$. This would mean, in principle,  that at the end of the collapse process. we would be left with states having a relatively  well defined value of  $|f|^2 $ but possibly different values of $f$.  We  do  not think this  will be  a problem,   because once  the unitary dynamics and the interactions  are taken into account, this kind of situation would be very unstable: any kind of unitary process which distinguished  $ f $ from  say $ -f $,  would lead to differences which would be subsequently distinguished by the collapse process.  In fact,  a more  realistic  analysis, where   backreaction  effects  would have to be  considered,  indicates  that   energetics  will   strongly    disfavor    field  configurations  with  large 
 spatio-temporal fluctuations in the phase. This follows from the fact  such  configurations  will  have  a relatively  large energy momentum tensor, and thus a  large spacetime curvature,   and as a consequence,   they will  be subjected to an increased  collapse rate. The   ensuing randomness in the dynamics  will only  decrease  when a  configuration   with  a rather  smooth $f$  is  arrived  at. This  is  analogous to the effect  considered  in \cite{Genesis}.   Thus  it  is  natural to expect that  ultimately the collapse will be to the $\hat{f}$  basis.

On the other   hand,  as  we  will be assuming  that the collapse rate increases  with curvature  in  an unbounded  fashion, we can expect that collapse effects will  accumulate   and dominate over  any dispersion in the high curvature region near to the black hole  singularity, more  precisely    as   the  quantum gravity region is approached, and thus we will assume that the collapse process leads  to a field eigenstate  on  hypersurfaces  that  are    close  enough to that region  as an idealization.

Finally, the particular choice $\hat{B}(x) = |\hat{f}(x)|^2$ can be justified by demonstrating that this reduces to the well established CSL model \cite{Pearle:89}, \cite{GRW:90}, in the non relativistic limit of a massive complex scalar field of mass $m$ (the non relativistic limit of a real scalar field or a massless field is less obvious). This can be seen from the well know correspondence
\begin{align}
\hat{f}(x) = \frac{e^{-imt}}{\sqrt{2m}} \hat{f}_{\rm nonrel}(x).
\end{align}
 The collapse basis is then
\begin{align}
\hat{B}(x) = |\hat{f}(x)|^2 = \frac{1}{2m}\hat{f}_{\rm nonrel}^{\dagger}(x) \hat{f}_{\rm nonrel}(x),
\end{align}
 where $\hat{f}_{\rm nonrel}^{\dagger}(x) \hat{f}_{\rm nonrel}(x)$ is the number density of non relativistic particles. Up 
 to a spatial smearing function, this is the collapse basis for the CSL model. The smearing is introduced either by spacetime 
 discreteness or the use of an auxiliary field to mediate the collapse process (see Appendix \ref{AX:AUX}).

\section{The setting}
 The  situation  we  want to consider  is that  corresponding to the formation of a black hole   by   the  gravitational collapse of  an initial matter   distribution  characterized  by a pure quantum state $ |\Psi_{0}\rangle $  describing a  relatively   localized  excitation of  the  field  $\hat  f $. 
  
   The  spacetime   is  supposed to be  described   by  a manifold  $M$   with a metric   $ g_{ab}$   defined    on $M$ except  for a  compact set $S_{QG}$ corresponding  to  the   region   where a   full quantum gravity  treatment   is required  and that is  taken to  just  surround the  location of  the  classical  singularity.
   This   characterizes the  formation and evaporation of an essentially   Schwarzschild   black hole,   supplemented by the  region $S_{QG}$ that  is not  susceptible   to a  metric  characterization and  where  a  full quantum theory of  gravity  is  needed to provide a suitable description.  We  assume that  $\partial S_{QG}$ is  a  compact   boundary   surrounding the   
quantum gravity region,    which,   by  assumption,  corresponds to that  region  where otherwise (i.e,  in the absence of   a  radical   modification of GR  due to QG  effects)  we would  
have encountered the black  hole singularity.     

  We  will  further make  some  relatively mild (and  rather common) assumptions   about  quantum   gravity.
  
   i) The first assumption, which we  have  already mentioned,  is that   QG  will cure  the    singularities  of  general  relativity,   however   in doing so   it will   require  that there   would   be   regions   where the  standard  metric  characterization of spacetime  does not apply.  This  is  what    in  our case  was  referred   as    the set $S_{QG}$
   
   ii)  We  will assume that Quantum  Gravity does not   lead   at the effective level to  dramatic  violations  of     basic    conservation laws  such as    energy or  momentum.
   
   iii) We  will assume that   the    spacetime region   that   results   at the other side of the QG region  is  a  reasonable   and  rather   simple  spacetime. 
   
With  these assumptions   we   can  already   make     some   simple predictions about the nature of the   full spacetime.

Given that by assumption the  effects of  the collapse dynamics  will be   strong only in the region with  high   curvature, and more explicitly in the regions   where the value of $ W^2$ ($R$ in the two dimensional models) is  large,   the   dynamics   characterizing the   early   evolution  of our initial pulse of matter will be   essentially the same as that found in  the  standard accounts of  black hole formation and   evaporation: The pulse   will   contract    due to   its  own  gravitational pull,  and  as  shown  by   Birkoff's  theorem the  exterior  region   will be  described  by  the Schwarzschild  metric;  the pulse  will  eventually  cross  the corresponding Schwarzschild  radius, and   generate  a Killing horizon   for the   exterior time-like Killing field $\xi^a$.   The   early  exterior  region  and   even the  region to the interior of the  Killing  horizon  but close to it  at early times  are regions of small curvature  and thus  the    picture   based on standard  quantum field theory in curved  spacetime  that leads to  Hawking radiation  will remain unchanged.   This by itself indicates that   essentially all the   initial   ADM mass of the spacetime  would  be radiated in the form of Hawking radiation   and will  reach  ${\cal  J}^+$ (asymptotic  null infinity).

  Next  let  us consider  the  spacetime  that emerges  at the other  side of the singularity.  Given   that essentially  all the   initial energy has been radiated to ${\cal  J}^+$   and in light of    assumptions ii)  above   the resulting spacetime  should  correspond to  one   associated with a  vanishing mass (this  would be the Bondi mass    corresponding to   a  spacetime hypersurface  lying   to the future of region $S_{QG}$  and  intersecting ${\cal  J}^+$  in  a  segment  to the future of that containing  the Hawking flux).   This  conclusion, together with assumption iii)  indicates that   this spacetime   region   should be  a simple  vacuum spacetime  which  we take for simplicity to correspond to  a  flat Minkowski region. 
  
Let  us  now  focus on the state of the quantum field $\hat f $.  The initial state,   as we  indicated,  
corresponds  to     the {\it in } vacuum   except for  a  pulse  of  matter   falling  under  its own  
gravity and  leading to the formation of a 
black hole.
 
  The   state can  be  represented  in the   first QFT  construction    in  section II  as:
 \begin{align} \label{initial state}
|\Psi_0\rangle    =    | Pulse \rangle^{in}_L  \otimes  | 0\rangle^{in}_R,
\end{align}
 where the    first  term    represents the  high  degree of excitation of  the few  modes  
 associated  with the   matter pulse   while the  second represents the  state of all  
 other modes  of the quantum field that  are  by  assumption in their   corresponding  
 vacuum state.    
   
As  is  well known  we can now  describe this  state in the quantization associated 
 with the  late  region,    on  which  every  Cauchy hypersurface   can  be  separated 
into the part  external  to the   Killing horizon  where  we   have the   approximate  
 Killing field  of the Schwarzschild  spacetime  and the  region interior to the Killing  
horizon where  one can use a  fiducial notion  ``particle'' to  define   creation and   
 anihilation operators  so that  the   in vacuum   state  can be  written as: 
\begin{align}
  | 0\rangle_{in} = {\cal N} \sum_F e^{-\beta_H E_F /2} |F\rangle^{int}\otimes |F\rangle^{ext},
\end{align}
   where  the  sum is  over the  sets   of   occupation number  for all modes $ F=\lbrace  F_{nj} \rbrace$ 
   (which  indicates that the  mode   $ n, j$  is  excited    by  $ F_{nj} $    quanta ),  $ E_F = \sum \omega_{nj} F_{nj} $ is the  total energy of the   state  according to the   notion  of    
   energy associated  with  the  asymptotic region, $ \beta_H $ is the Hawking   
   thermal   coefficient,  and ${\cal  N}$   is  a normalization   constant.
   
    At these  late  times the   excitations  associated  with  the  in-falling pulse  are  all located   in the  region  interior   to the Killing Horizon  so that   we can   write the state (\ref{initial state})  simply as: 
 	\begin{align}
 |\Psi_0\rangle  = {\cal N} \sum_F e^{-\beta_H E_F /2}   ( | Pulse \rangle^{in}_L \otimes  |F\rangle^{int}_R ) \otimes |F\rangle^{ext}_R
\end{align}
   where the part in parenthesis   corresponds to the    black hole  interior  region  and   the rest  to the  exterior. 
  
  The point of   writing things in this manner is to   underscore the fact that  both    the  collapse  dynamics  and  the  changes in the state  associated  with  quantum gravity will  only  affect  the  modes  in the  black hole  interior  region. In the case of the collapse  dynamics this   follows from the assumption  that the collapse parameter  $ \gamma$  is strongly  dependent of curvature and thus its effects   will only be relevant  in regions of high curvature.


\section{Collapse dynamics in the CGHS black hole}
  As  we   have     explained, one of the    assumptions  that    underlies   the   present  approach to deal  with  the information   question  during  the Hawking evaporation of the black hole  is  that the  collapse  dynamics,  although    valid  everywhere,  deviate  most strongly  from the     unitary   evolution of  standard  quantum theory in  the regions  where   curvature becomes   large.  In   the    two dimensional context, this  is achieved  by assuming the   that     parameter   $\gamma$ controlling the  strength of the modifications   is a function of  the scalar  curvature $R$. Thus  the   changes to the  quantum state  of the system result mainly   from the    nontrivial evolution  occurring in the  region  interior  to the  black hole   horizon,  and to the future of the   matter   shell.  For  simplicity we   will therefore ignore the modification of  the quantum  state of the field  resulting from the   dynamics  in the exterior region   and    the  flat region   before the  matter shell and  focus only  in the effects of the     
 collapse dynamics in the interior of the black hole  lying to the  future.  We call this    {\it the collapse region}.

\begin{figure}
\centering
\includegraphics[scale=0.3]{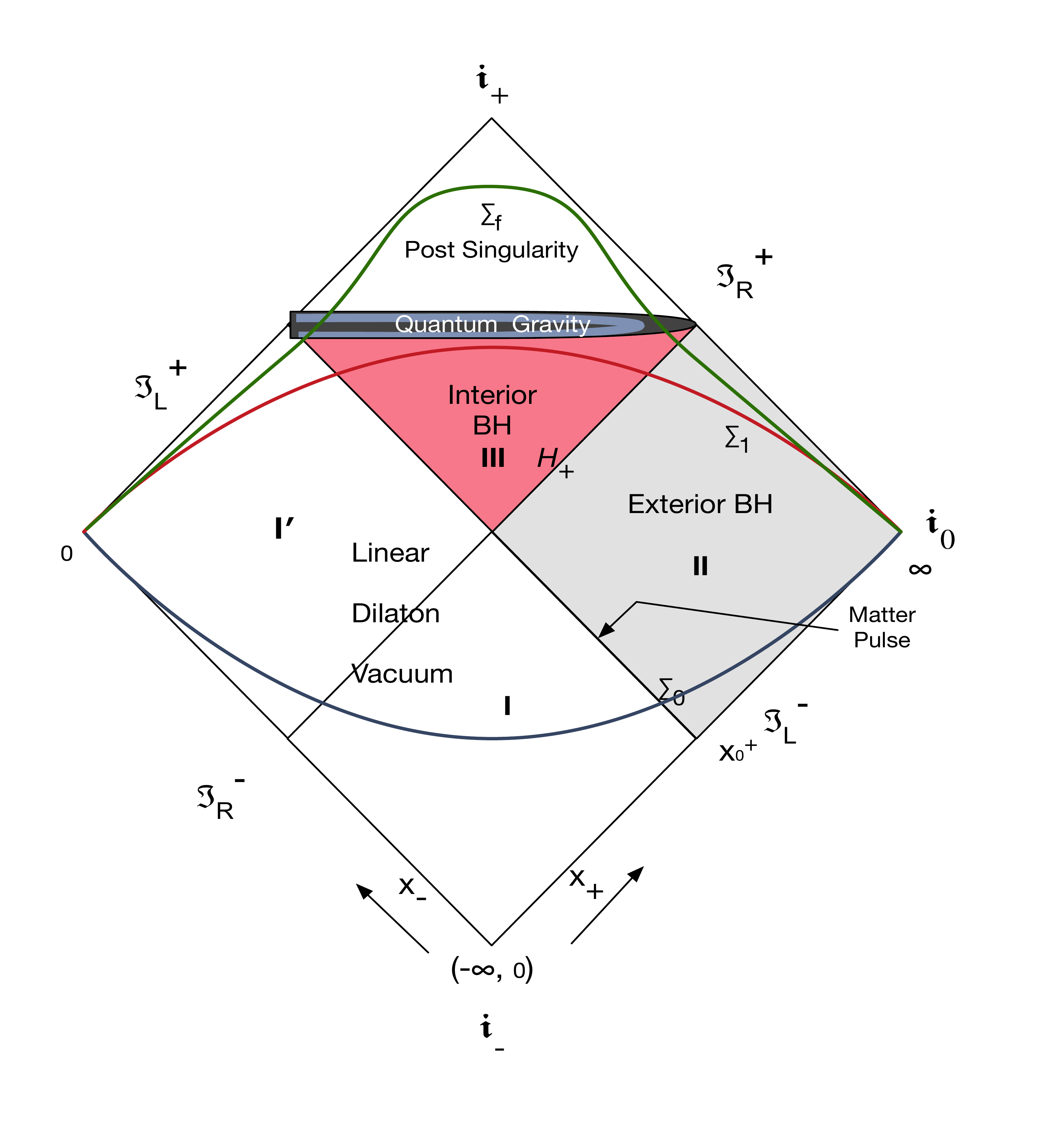}
\caption{Penrose diagram for CGHS spacetime including the post- quantum gravity   empty region. }
\label{cghs}
\end{figure}

      With  these considerations  we take the initial state  at a hypersurface   lying   well  to  the past    of the    collapse region 
      ( for instance   on $\Sigma_0$ in Fig. \ref{cghs}). This can  be expressed in terms of the corresponding  density 
   matrix    as: 
\begin{align}
\rho  (\Sigma_0) = {\cal N}^2 \sum_{FG} e^{-\beta_H (E_F+E_G)/2 } |F\rangle^{int}\langle G|^{int} \otimes |F\rangle^{ext} \langle G|^{ext}.
\end{align}
 We  can now    simplify  things   using the   basis  of  eigen-states of  collapse  operators    which we   will refer to as   {\it the  collapse basis}.  We  thus rewrite the  above   density matrix  in the form  
\begin{align}
\rho  (\Sigma_0) = {\cal N}^2 \sum_{ij}\sum_{FG} e^{-\beta_H (E_F+E_G)/2 } \langle G|f_j\rangle^{int} \langle f_i |F\rangle^{int}  |f_i \rangle^{int} \langle f_j|^{int} \otimes |F\rangle^{ext} \langle G|^{ext}.
\end{align}

 Next  we  use the fact that,   in the collapse  region,    especially in the   late  part thereof,  the    collapse  dynamics    becomes  extremely strong and  effective and thus  drives the  state of the system to  one of  the   eigen-states of the collapse operators. This  allows  us to  write   the   state  representing  an ensemble of systems      initially prepared in the  same   state (\ref{initial state}),    at  any  hypersurface  $\Sigma_1$  lying     just  before the  would-be  classical singularity --or more precisely the  quantum  gravity region--(see Fig. \ref{cghs}),   after the  complete collapse process   has  taken place  as,  
\begin{align}
\rho (\Sigma_1) &= {\cal N}^2 \sum_{i}\sum_{FG} e^{-\beta_H (E_F+E_G)/2 } \langle G|f_i\rangle^{int} \langle f_i |F\rangle^{int}  |f_i \rangle^{int} \langle f_i|^{int} \otimes |F\rangle^{ext} \langle G|^{ext}.
\end{align}
Finally,  we   need to consider  the    system as  it emerges on the  other side of the quantum gravity region, i.e  the  state   describing the     ensemble   after  the     would-be  classical singularity. 
As  we have  discussed in the introduction   we    assume that    quantum gravity  
would resolve the singularity and lead  on the other side  of it,  to some  reasonable 
spacetime  and state of  the quantum fields. We  now    consider    
the characterization of the system  on  a hypersurface  lying just to the future of  the   
would-be  classical singularity.  Such  a hypersurface   would  not  be a Cauchy  
hupersurface  as it would  intersect    ${\cal J}^+$   rather than $i^0$.   As  such   one   
can  partially characterize   the  state of fields  on  it  by    the value of the Bondi mass.   
It  is clear,   as  have argued  in the   the introduction,  that if we assume that Quantum 
Gravity   does not lead  to large violations of   energy and  momentum conservation 
laws,   the only possible  value for this  Bondi mass  would have to be  the mass of the 
initial   matter shell  minus the energy emitted   as Hawking   radiation,  which is  
present to the   past of the singularity on ${\cal J}^+$. This  remaining mass  will  thus have to be   
very small.  

The task for quantum gravity is to turn the internal state, post singularity into a straightforward low energy state. For simplicity assume that it is the vacuum
\begin{align}\label{Post-singularity-vacuum}
|f_i\rangle^{int}|Pulse\rangle_L \rightarrow |0^{post-sing}\rangle,
\end{align}
the particular $|f_i\rangle^{int}$ being chosen by the collapse process. 

This means that the final state characterizing  the  ensemble of systems (on $\Sigma_f$)    should be of the form: 
\begin{align}
\rho_{final} &= {\cal N}^2\sum_{i}\sum_{FG} e^{-\beta_H (E_F+E_G)/2 } \langle G|f_i\rangle^{int} \langle f_i |F\rangle^{int}  |0^{post-sing} \rangle \langle 0^{post-sing}| \otimes |F\rangle^{ext} \langle G|^{ext}
\nonumber\\
&={\cal N}^2\sum_{FG} e^{-\beta_H (E_F+E_G)/2 } \langle G|F\rangle^{int}  |0^{post-sing} \rangle \langle 0^{post-sing}| \otimes |F\rangle^{ext} \langle G|^{ext}
\nonumber\\
&={\cal N}^2\sum_{F} e^{-\beta_H E_F }  |0^{post-sing} \rangle \langle 0^{post-sing}| \otimes |F\rangle^{ext} \langle F|^{ext}. \nonumber\\
&= |0^{post-sing} \rangle \langle 0^{post-sing}| \otimes  \rho_{thermal}^{ext}
\end{align}
That  is,   the    system has   evolved   from  an initially pure state  to  state representing   the proper thermal state of radiation  on  the
 early part of ${\cal J}^+$ 
 and the vacuum state  afterwards.

\section{Energy  considerations}

 One  of the most  serious challenges  one   faces  when  attempting     to construct   relativistic models of   spontaneous   dynamical   reduction of the wave function, either of  the    discrete or  continuous  kind,   
is  their  intrinsic    tendency   to  predict    the violation  of energy   conservation by  infinite   amounts: 
The problem  is resolved   in the non-relativistic  setting where  one can easily control the magnitude of that kind of effect,  by  relying on suitable   spatial  smearings of the   collapse operators,   usually taken to be the   position operators  for the individual particles that make  up the system.

 When passing to a relativistic   context  the  tendency is for   energy  violation to  become  unbounded  unless special care is  used  in the  construction of the theory  to  ensure it does not. 

     This  issue  becomes relevant in the present  context  at two places. First and foremost   at the point   where one  wants  to  consider the back reaction    of the spacetime  metric  to the   changes   in the quantum state of the field  $\hat f$  induced  by the collapse dynamics.   The second place  where the issue  appears is the point  where   one     considers   the role of   the quantum gravity region.  In a previous treatment the  argument was that, provided   that  quantum gravity did not result in large violations of energy conservation  one can  expect the  state  after the   quantum gravity region  to   correspond energetically   to  the   content of the  region just before the    would be singularity, and that  this region    would have almost  vanishing  energy content   being   made  up of  the  positive   energy contribution of the    collapsing matter shell and the   negative energy contribution of the   in-falling   counterpart  to the Hawking flux. 
We  would face a serious  problem  with this  argument if  the    region just before the    would-be singularity   could  contain  an   arbitrarily large  amount of  energy   as  a result of the   
unboundedness of the  violation of   energy  conservation brought about by the collapse dynamics.
 In that case  we  would not be  able to reasonably argue  for the step (\ref{Post-singularity-vacuum}).

There are various  schemes  whereby  this issue  can be   tackled:

1)  We  might  consider a fundamental discreteness of spacetime (which however  as  discussed in\cite{LIV} should  not  be tied to violations of  special relativity).

2)  We might  adjust  the choice  of   collapse operators and    provide  a   sensible   spacetime   smearing scheme  for them that relies on  the  energy momentum of the matter fields  or  on the  geometric  structure of   the    curved  spacetime.  In this    context is  it worth   noting that   when   one   considers  that  the   parameter  controlling the   strength  or  intrinsic   rate   of the   collapse dynamics   depends on the  spatial curvature   (i.e. the  Ricci scalar  $ R$  in  2 dimensions    and  something like the  Weyl   tensor,    through  $W^2  =  W_{abcd}W^{abcd}$,  in  the   more realistic  4  dimensional case)  one  might assume that   in flat space-times the   collapse rate   actually vanishes    removing  most   concerns  about    the stability of the vacuum  in these theories.   In that case one  would adopt the position that the collapse  associated with  individual particles  in  the non-relativistic  quantum mechanical   context  is  actually derived  from the  small   deformation of flat spacetime associated  to that  same  particle. That  is,  one  would consider that   the particle's  energy momentum      curves the spacetime  and this in turn  turns-on  the  quantum collapse    dynamics.  
 This   is  only a  rough   idea  at this point  but one   that certainly  seems  worthy of further exploration.  

3)  We  might rely on  the  effective  smearing  provided  by  the  use of  the    auxiliary pointer field   as    a   way to      introduce  the smearing    procedure without   seriously affecting the    simplicity of the   treatment  as    discussed in the Appendix \ref{AX:AUX}   below.

\section{Conclusions}

We  have   studied  the possibility   of  accounting for the information  loss  in the    processes  of formation and    Hawking evaporation   of  a black    hole  through  the  explicit use  of  a relativistic version of  a  dynamical reduction theory.     In previous    works it has  been argued that  the  consideration of  theories  involving   departure from   the standard    Schr\"odinger unitary dynamics  offers a promising path to   dealing  with what  many  researchers in the community   considered  as  one of the  most  challenging  paradoxes  of modern theoretical physics.     Those   works    were     carried out using a non-relativistic  version of  dynamical reduction theories   known as   Continuous  Spontaneous Localization, and one  of the   open issues\footnote{ The other major limitation  that those initial treatments  referred to the question of back reaction     of the spacetime to the quantum   aspects of   the matter. That  issue  is the topic of a coming paper  by some of us\cite{RST-Us} } in   those treatments was  whether   similar results  could  be obtained  relying on fully relativistic  settings.

  The  present work   provides a  positive   answer  in the form of  proof   of  existence of a  relativistic  approach    that   leads  essentially to the same  results   as those  of   the previous   non-relativistic  treatments. However it is  clear that we are not yet  in the possession of  a  fully  satisfactory   scheme.
 
 For   that     we  need to consider in detail the  issues   related to  energy  production and  its  
 possible     back  reaction  effects.    Furthermore   eventually  one   would like to  consider     the   
 issue  of uniqueness   and  completeness  in   the sense of determining   the collapse operators  valid 
 for  a general setting   that reduces  to  the   
appropriate ones (i.e   smeared  particle position operators)  in  the   non-relativistic  situations (the ones  treated by the   standard CSL or GRW  theories),   and  finding the   dependence of  the parameters  such as $\gamma$ on  the   spacetime curvature.

\section{Acknowledgements}
DB is supported by the Templeton World Charity Foundation. SKM is an International Research Fellow of the Japan Society for the Promotion of Science. DS  acknowledges partial financial support from DGAPA-UNAM project IG100316 and  by CONACyT project 101712.

\appendix
\section{Foliation independence of collapse process}
\label{Ax:folindep}
Consider two collapse events occurring at spacelike separated points $x$ and $y$ with $\Sigma_i\prec x,y\prec \Sigma_f$ (meaning that the points $x$ and $y$ are not to the past of $\Sigma_i$ and not to the future of $\Sigma_f$). An explicit choice of foliation places $x$ and $y$ in a sequence. Suppose that $x$ occurs first on surface $\Sigma_1$ and $y$ occurs second on surface $\Sigma_2$ with $\Sigma_i\prec\Sigma_1\prec\Sigma_2\prec\Sigma_f$. We therefore have
\begin{align}
|\Psi_{\Sigma_f} \rangle = \hat{U}[\Sigma_f,\Sigma_2]\hat{L}_y(Z_y)\hat{U}[\Sigma_2,\Sigma_1]\hat{L}_x(Z_x)\hat{U}[\Sigma_1,\Sigma_i]|\Psi_{\Sigma_i}\rangle,
\end{align}
with 
\begin{align}
\mathbb{P}\left(Z_x,Z_y\right||\Psi_{\Sigma_i}\rangle) 
 = \frac{\langle \Psi_{\Sigma_f} |\Psi_{\Sigma_f}\rangle }{\langle \Psi_{\Sigma_i}|\Psi_{\Sigma_i}\rangle},
\label{PROB2a}
\end{align}
which follows from (\ref{PROB})  by making use of the definition of conditional probability.
 Now suppose instead that we choose an alternate foliation in which the collapse event at $y$ occurs first on surface $\Sigma_1'$ and $x$ occurs second on surface $\Sigma_2'$ with $\Sigma_i\prec\Sigma_1'\prec\Sigma_2'\prec\Sigma_f$. Now 
\begin{align}
|\Psi'_{\Sigma_f} \rangle = \hat{U}[\Sigma_f,\Sigma_2']\hat{L}_x(Z_x)\hat{U}[\Sigma_2',\Sigma_1']\hat{L}_y(Z_y)\hat{U}[\Sigma_1',\Sigma_i]|\Psi_{\Sigma_i}\rangle,
\end{align}
with 
\begin{align}
\mathbb{P}'\left(Z_x,Z_y\right||\Psi_{\Sigma_i}\rangle) 
 = \frac{\langle \Psi'_{\Sigma_f} |\Psi'_{\Sigma_f}\rangle }{\langle \Psi_{\Sigma_i}|\Psi_{\Sigma_i}\rangle}.
\label{PROB2b}
\end{align}
We now show that given the conditions
\begin{align}
[\hat{L}_x(Z_x),\hat{L}_y(Z_y)]= 0,
\label{ACOM2}
\end{align}
and
\begin{align}
[\hat{L}_x(Z_x),\hat{\cal H}_{\rm int}(y)]= 0,
\label{ACOM3}
\end{align}
for spacelike separated $x$ and $y$, then $|\Psi'_{\Sigma_f} \rangle = |\Psi_{\Sigma_f} \rangle$. In order to do this we define a surface $\Sigma_1\prec\Sigma_{xy}\prec\Sigma_2$ on which both points $x$ and $y$ are found. We can then write
\begin{align}
|\Psi_{\Sigma_f} \rangle &= \hat{U}[\Sigma_f,\Sigma_2]\hat{L}_y(Z_y)\hat{U}[\Sigma_2,\Sigma_{xy}]\hat{U}[\Sigma_{xy},\Sigma_1]\hat{L}_x(Z_x)\hat{U}[\Sigma_1,\Sigma_i]|\Psi_{\Sigma_i}\rangle \nonumber\\
&= \hat{U}[\Sigma_f,\Sigma_2]\hat{U}[\Sigma_2,\Sigma_{xy}]\hat{L}_y(Z_y)\hat{L}_x(Z_x)\hat{U}[\Sigma_{xy},\Sigma_1] \hat{U}[\Sigma_1,\Sigma_i]|\Psi_{\Sigma_i}\rangle \nonumber\\
 &=\hat{U}[\Sigma_f,\Sigma_{xy}]\hat{L}_x(Z_x)\hat{L}_y(Z_y)\hat{U}[\Sigma_{xy},\Sigma_i]|\Psi_{\Sigma_i}\rangle.
\end{align}
The second line uses the fact that 
\begin{align}
\left[\hat{L}_x(Z_x),\hat{U}[\Sigma,\Sigma']\right] = 0,
\end{align}
if $x$ is found on both $\Sigma$ and $\Sigma'$, which follows from (\ref{ACOM3}). The third line follows from (\ref{ACOM2}). Next we define the surface $\Sigma'_{xy}$ on which both points $x$ and $y$ are found and such that $\Sigma'_1\prec\Sigma'_{xy}\prec\Sigma'_2$, along with a further surface $\Sigma''_{xy}$, also containing $x$ and $y$ and satisfying $\Sigma_i\prec\Sigma''_{xy}$; $\Sigma''_{xy}\prec\Sigma_{xy}$; and $\Sigma''_{xy}\prec\Sigma'_{xy}$. We then have
\begin{align}
|\Psi_{\Sigma_f} \rangle &= \hat{U}[\Sigma_f,\Sigma_{xy}]\hat{L}_x(Z_x)\hat{L}_y(Z_y)\hat{U}[\Sigma_{xy},\Sigma''_{xy}]\hat{U}[\Sigma''_{xy},\Sigma_i]|\Psi_{\Sigma_i}\rangle \nonumber\\
&= \hat{U}[\Sigma_f,\Sigma''_{xy}]\hat{L}_x(Z_x)\hat{L}_y(Z_y)\hat{U}[\Sigma''_{xy},\Sigma_i]|\Psi_{\Sigma_i}\rangle \nonumber\\
&= \hat{U}[\Sigma_f,\Sigma'_{xy}]\hat{U}[\Sigma'_{xy},\Sigma''_{xy}] \hat{L}_x(Z_x)\hat{L}_y(Z_y)\hat{U}[\Sigma''_{xy},\Sigma_i]|\Psi_{\Sigma_i} \rangle \nonumber\\
&= \hat{U}[\Sigma_f,\Sigma'_{xy}]\hat{L}_x(Z_x)\hat{L}_y(Z_y)\hat{U}[\Sigma'_{xy},\Sigma_i]|\Psi_{\Sigma_i}\rangle \nonumber\\
&= \hat{U}[\Sigma_f,\Sigma'_2]\hat{U}[\Sigma'_2,\Sigma'_{xy}]\hat{L}_x(Z_x)\hat{L}_y(Z_y)\hat{U}[\Sigma'_{xy},\Sigma'_1] \hat{U}[\Sigma'_1,\Sigma_i]|\Psi_{\Sigma_i}\rangle \nonumber\\
&= \hat{U}[\Sigma_f,\Sigma'_2]\hat{L}_x(Z_x)\hat{U}[\Sigma'_2,\Sigma'_{xy}]\hat{U}[\Sigma'_{xy},\Sigma'_1]\hat{L}_y(Z_y)\hat{U}[\Sigma'_1,\Sigma_i]|\Psi_{\Sigma_i}\rangle \nonumber\\ 
&= |\Psi'_{\Sigma_f}\rangle.
\end{align}
Using this result it follows from (\ref{PROB2a}) and (\ref{PROB2b}) that the probability density for the pair of collapse outcomes $Z_x$ and $Z_y$ is independent of the choice of foliation. Iteration of the above procedure for further collapses demonstrates foliation independence of the complete set of collapse outcomes $\{Z_{x_j}|\Sigma_f \succ x_j\succ \Sigma_i\}$ occurring at the set of collapse locations $\{x_j|\Sigma_f\succ x_j\succ\Sigma_i\}$ between any $\Sigma_i$ and $\Sigma_f$.

\section{Use of an auxiliary field}
\label{AX:AUX}

A way to understand the infinite energy increase of the collapse dynamics described in section \ref{CovColl} is to notice that each collapse on the quantum state  occurs at a single point on  the spacetime. This results in sharp  spatio-temporal  discontinuities  in the  state of the  field,  and hence a large energy increase. In order to prevent this the collapse should happen smoothly. For a quantum field this means that whenever a local collapse occurs it should act over some spacetime region rather than at an infinitesimal space time point. This requires some form of smeared interaction. In order to facilitate this we use a new type of relativistic quantum field which we call the {\it pointer field} as introduced in \cite{BED1,BED2}. This field has an independent degree of freedom at each space time point.  We  will denote it by  $\hat\psi$  (not to be  confused  with the matter field $ \hat f$).

The commutation properties of the pointer field are as follows:
\begin{align}
\left[\hat\psi(x),\hat\psi^{\dagger}(x')\right] = \frac{1}{\sqrt{g(x)}}\delta^4(x-x'); \quad
\left[\hat\psi(x),\hat\psi(x')\right] = 0.
\end{align}
Notice that the Dirac delta extends over the whole space time, not just over a hyper surface. Given these annihilation and creation operators we can define a smeared field operator
\begin{align}
\hat{A}(x) = \int d^4y\sqrt{g(y)} s_A(x,y) \left[ \hat\psi(y) + \hat\psi^{\dagger}(y)\right].
\end{align}
and a smeared number density operator which will be the collapse basis of equation (\ref{L})
\begin{align}
\hat{B}(x) = \int d^4y\sqrt{g(y)} s_B(x,y) \hat\psi^{\dagger}(y)\hat\psi(y),
\end{align}
The smearing functions $s_A$ and $s_B$ are assumed to be defined in terms of the (fixed) space time properties such as local curvature. They should each satisfy certain properties of smoothness and finiteness under integration to be determined by their consequences for the relativistic collapse theory. 

We assume that there is an interaction between ordinary matter fields and the pointer field of the form 
\begin{align}
\hat{\cal H}_{\rm int}(x) = \nu |\hat{f}(x)|^2 \hat{A}(x).
\end{align}
where $\nu$ is a coupling parameter, which, as we will see below, affects the effective collapse rate of the matter field. The state $|\Psi_{\Sigma}\rangle$ now includes the state of both matter field $\hat{f}$ and the pointer field $\hat\psi$. The required micro causality conditions (\ref{COM2}) and (\ref{COM3}) can be used to begin to constrain the form of $s_A$ and $s_B$. In general it follows from the above commutation properties that
\begin{align}
\left[\hat{A}(x),\hat{A}(x')\right] = 0; \quad \left[\hat{B}(x),\hat{B}(x')\right] = 0,
\label{COM0}
\end{align}
for all space time points $x$ and $x'$. It also follows that for space like separated $x$ and $x'$,
\begin{align}
\left[\hat{A}(x),\hat{B}(x')\right] = 0,
\label{COM}
\end{align}
provided that the domain of $s_A(x,y)$ only includes points where $y$ is inside the future light cone of $x$, and the domain of $s_B(x,y)$ only includes points where $y$ is inside the past light cone of $x$. This commutation property result in (\ref{COM3}).

As  a concrete proposal  for  $s_A$ on  a  spacetime  with metric  $g_{ab} $
   we    will  take:
 \begin{align}\label{eq:sa}
 s_A(x,y)  =   \Theta_{I^+}(x, y)  \times   e^{- ( \beta  \int_{c(x,y)} B_{abcd} T^a T^b T^c T^d  d\tau )^n},
\end{align}
where  the integration  measure  $d\tau$ along the geodesic  is the differential of  $\tau $ the  invariant line element, rather than volume element
  where  $n$ is  some   positive  integer,  $\beta$ is a   suitably chosen  dimensional parameter,  $ \Theta_{I^+}(x, y) $  is  the characteristic  function of  $I^+ $, the  chronological  future of  $x$,  that is 
  $ \Theta_{I^+}(x, y)  = 1$  iff  $ y	\in  I^{+} (x) $ and  vanishes otherwise, 
  $ c(x,y) $  is the  causal   geodesic   connecting  $x$  and  $y$   (which  we  will  assumed to be unique \footnote{This of course will hold  for  $y$  in a   convex  normal  neighborhood   of $x$. For  points  outside this  region we can   replace the prescription  to one  where   we   replace   the integral    along  a single geodesic  by  the  sum  of   integrals   over all such  geodesics.}),   $T^a$ is the   tangent to  the geodesic   $c$   by proper time, and   $B_{abcd} $   is the  Bell  tensor of the  spacetime  metric $g_{ab}$.
  
  We  note that   as  the   $T^a$  are  future directed time-like   vectors  the  integrand  in  the  above  equation   is  positive semi-definite.  That is  $ B_{abcd} T^a T^b T^c T^d \geq 0$.    In fact,    
  generically  this quantity will vanish only along the  principal  null  directions of the  Weyl  tensor,  which 
  as we know \cite{Bonilla-Senovilla},    are  in  general  just a  discrete set of directions  (4 in  4  spacetime dimensions).  It    
   would be only  when such  null directions coincide   with  a   tangent  $T^a$ along  the full  null  geodesic  connecting $x$  and  $\tilde y \in J^{+}$  that such an integral  would  vanish.  It is  therefore,  only in  
  those  very   unusual cases (where  such a $ \tilde  y $  exists),   that   for points $y\in I^{+} (x)$   
  that  approach  the  points $\tilde y$, that  the   
  integral $ \int_{c(x,y)} B_{abcd} T^a T^b T^c T^d d\tau $ might fail to be  bounded  from  below  by a positive   number.    Otherwise,  in the  generic  situations,   the    functions $  s_A(x,y)$  will    rapidly decrease   as  the point  $y$    gets    further from $x$  even along the   directions that approach those in the  ``null  cone"  corresponding to the boundary  of the chronological  future  of  $x$: $\partial I^{+} (x) $.   
  
  Analogously we   can set, 
   \begin{align}
 s_B(x,y)  =   \Theta_{I^-}(x, y)  \times   e^{-(\beta  \int_{c(x,y)} B_{abcd} T^a T^b T^c T^d d\tau )^n} ,
\end{align}
  where  $ \Theta_{I^-}(x, y) $  is  the characteristic  function of  $I^- $, the  chronological  past of  $x$,  that is   $ \Theta_{I^-}(x, y)  = 1$  iff  $ y	\in  I^{-} (x) $ and  vanishes otherwise.

To understand how the collapse mechanism works consider first the interaction term $|\hat{f}(x)|^2\hat{A}(x)$. This has the effect of coupling the state of the $\hat{f}$ field to the state of the $\hat\psi$ field. An excited matter field will lead to an excitation of the pointer field in the local region determined by $s_A$. For a matter field in a superposition of different $\hat{f}$ states this interaction will lead to an entanglement between the different $\hat{f}$ states and different states of the pointer field. 

The state  of the  pointer field  is the one that is now  subjected  directly to  the collapse  dynamics  of Sction III. The   state of the system is   an  element of the   product  space between the   Hilbert space of the  matter fields  and that of the pointer field. 

The action of the collapse operator leads to a collapse of the pointer field in the (smeared) number density basis. They act as quasi projectors onto a number density state with a central value determined by the random choice $Z_x$. This causes the pointer field to collapse towards a state of definite number density. Since the matter field is entangled with the pointer field, the collapse of the pointer field induces a collapse of the matter field state in the field state basis.

Now  we   derive an effective characterization of  the  resulting   collapse dynamics  involving just the  quantum field  $\hat f$. Starting with the fully covariant model we will trace out the the pointer field leaving an approximate dynamical equation for the  $\hat{f}$ state. We start by choosing a particular foliation parametrized by a time coordinate $t$,  and  write the spacetime metric   in terms of the standard  3+1  decomposition as:
\begin{align}
dS^2 = -  ( N^2 - h_{ij}  N ^i N^j ) dt^2  - 2  h_{ij} N^j dx^i dt  +  h_{ij}dx^i dx^j,
\end{align}
where  $ N$   is the  lapse  function  and  $ N^i $ are  the components of the  shift   vector   characterizing the foliation  and  $h_{ij} $ are  the components  of the induced  metric  on  the corresponding spatial  hyper surface. 
 
The equation describing the interaction between the scalar field and the pointer field is then
\begin{align}
\frac{d}{dt}|\Psi_{t}\rangle = -i  \nu \int d^3 x N\sqrt{h} |\hat{f}(x)|^2\hat{A}(x) |\Psi_{t}\rangle,
\end{align}
where for now we ignore the collapses. The state is always pure and so the density matrix can be written as
\begin{align}
\hat\rho_t = |\Psi_{t}\rangle \langle \Psi_{t}|.
\end{align}
We further assume that the pointer field is always in an approximate vacuum state\footnote{ Recall  that the  
auxiliary  field  is  not a standard  quantum field.  In  particular  the   corresponding   pointer vacuum state is defined by  $\psi(x)|0\rangle = 0$. }
 so that we can write
\begin{align}
\hat\rho_t= \hat\rho^{f}_t \otimes \hat\rho^{\psi}_t.
\end{align}
This assumption requires that the coupling parameter $\nu$ is weak and that the pointer field collapses are able to resolve very small differences in number density so that the collapse occurs even after a very weak interaction between $\hat{f}$ and $\hat\psi$. This requires choosing $\zeta$ to be sufficiently small. The master equation describing the development of the density matrix is
\begin{align}
\frac{d}{dt}\hat\rho_t = -i\left[\hat{H}_t, \hat\rho_t\right],
\end{align}
where the interaction Hamiltonian is given by
\begin{align}
\hat{H} =  \nu\int d^3 x N\sqrt{h}|\hat{f}(x)|^2\hat{A}(x).
\end{align}
We can use the Born approximation, valid for weak interactions, to find an approximate solution to the master equation
\begin{align}
\hat\rho_t \simeq \hat\rho_0 - i\int_0^t dt' \left[\hat{H}_{t'},\hat\rho_t\right].
\end{align}
This can then be inserted back into the master equation
\begin{align}
\frac{d}{dt}\hat\rho_t \simeq -i\left[\hat{H}_t, \hat\rho_0\right] - \int_0^t dt' \left[\hat{H}_t,\left[\hat{H}_{t'},\hat\rho_t\right]\right].
\end{align}
Next we take the partial trace over the pointer field degrees of freedom. To do this we use
\begin{align}
{\rm Tr}_{\psi} \left[\hat{H}_t,\hat\rho_0\right] = \nu
\int d^3 x N\sqrt{h} \left[|\hat{f}(x)|^2,\hat\rho^{f}_0\right]
{\rm Tr}_{\psi}\left[\hat{A}(x)\hat\rho^{A}_0\right],
\end{align}
and
\begin{align}
{\rm Tr}_{\psi} \left[\hat{H}_t,\left[\hat{H}_{t'},\hat\rho_t\right]\right] = \nu^2
\int d^3 x N\sqrt{h}\int d^3 x' N'\sqrt{h'} \left[|\hat{f}(x)|^2,\left[|\hat{f}(x')|^2,\hat\rho^{f}_t\right]\right]
{\rm Tr}_{\psi}\left[\hat{A}(x)\hat{A}(x')\hat\rho^{A}_t\right].
\end{align}
Now assume that the function $s_A(x,y)$ can be reasonably well approximated by a delta function
\begin{align}
s_A(x,y) = \frac{\eta(x)}{\sqrt{g(x)}}\delta^4(x-y),
\label{fchoice}
\end{align}
where $\eta(x)$ is a spacetime dependent scale factor.
Together with the assumption that the pointer field density matrix is approximately vacuum this results in
\begin{align}
{\rm Tr}_{\psi}\left[\hat{A}(x)\hat\rho^{\psi}_0\right] = 0,
\end{align}
and
\begin{align}
{\rm Tr}_{\psi}\left[\hat{A}(x)\hat{A}(x')\hat\rho^{\psi}_t\right] = \frac{\eta^2(x)}{\sqrt{g(x)}}\delta^4(x-x').
\end{align}
Putting all this together we find the master equation for the scalar field to be of the form
\begin{align}
\frac{d}{dt}\hat\rho^{f}_t \simeq  - \nu^2 \int d^3 xN\sqrt{h} \;\eta^2(x)
\left[|\hat{f}(x)|^2,\left[|\hat{f}(x)|^2,\hat\rho^{f}_t\right]\right].
\label{masterXi}
\end{align}
This is of precisely the same form as equation (\ref{PHICOLL}) once we identify $\gamma \equiv  \mu^2/  8\zeta$  with $\nu^2 \eta^2$. This form of the master equation predicts infinite energy increase. This is tempered by choosing a form for $s_A(x,y)$ which is not a delta function and (\ref{masterXi}) should be considered an idealized limit for describing collapse.
 
In Section \ref{CovColl} we demonstrated that a relativistic master equation of this form reduces to the non relativistic CSL model. The correspondence can be made more precise by choosing a suitable frame of reference defined by the coordinates ${\bf x},t$ in which $s_A(x,y)$ [invariantly defined according to (\ref{eq:sa})] takes the approximate form, 
\begin{align}
s_A(x,x') \simeq \delta(t-t') \bar{s}({\bf x} - {\bf x}') 
\end{align}
where  we  assume that  we  have tuned the   choice of the parameter $\beta$   so  as  to ensure that, in   ordinary  laboratory situations,  where the spacetime  is  almost  flat (except  for the  curvature induced  by the  few  particles  involved), $\bar{s}$  reduces  approximately  to the standard CSL smearing function.
\begin{align}
\bar{s}({\bf x}) = \left(\frac{\alpha}{2\pi}\right)^{\frac{3}{2}}\exp\left(-\frac{\alpha}{2} {\bf x}^2\right),
\end{align}
with $1/\sqrt{\alpha}$ the GRW length scale. The resulting non relativistic CSL model is well known and produces finite energy increase which can be kept suitably small by appropriate choice of the parameters.


\begin{thebibliography}{99}

\bibitem{hawk} S. W. Hawking, ``Particle Creation By Black Holes," Commun. Math. Phys. {\bf 43}, 199 (1975) [Erratum-ibid. 46, 206 (1976)].

\bibitem{hawk2} S. W. Hawking, ``Breakdown of Predictability in Gravitational Collapse," Phys. Rev. D {\bf 14}, 2460 (1976).

\bibitem{Mathur1} 
  S.~D.~Mathur,
   ``The Information paradox: A Pedagogical introduction,''
  Class.\ Quant.\ Grav.\  {\bf 26}, 224001 (2009)
  [arXiv:0909.1038 [hep-th]].
	


\bibitem{Okon2} E.  Okon \& D.  Sudarsky, ``The Black Hole Information Paradox and the Collapse of the Wave Function,"   {\it Found. of Phys.} {\bf 45}, 461-470 (2015).


\bibitem{bh-singularity} A. Ashtekar and M. Bojowald, ``Quantum geometry and the Schwarzschild singularity,'' Class. Quant. Grav. \textbf{23}, 391-411 (2006).	

    
      
\bibitem{sGidd92}
      S. B. Giddings,
      ``Quantum mechanics of black holes,"
      [arXiv:hep-th/9412138v1].

\bibitem{aStr95}
      A. Strominger, ``Les Houches Lectures on Black Holes,"
      [arXiv:hep-th/9501071v1].

\bibitem{Benachenhou:1994af}
  F.~Benachenhou,
  ``Black hole evaporation: A Survey,''
  [hep-th/9412189].

      
\bibitem{ST} L. Susskind and L. Thorlacius, ``Hawking radiation and back-reaction," Nucl. Phys. B {\bf 382}, 123-147 (1992).

\bibitem{RST} J. G. Russo,  L. Susskind and L. Thorlacius, ``The endpoint of Hawking radiation,'' Phys. Rev. D {\bf 46}, 3444 (1992).

\bibitem{COMPL} 	 L. Susskind, L. Thorlacius and J. Uglum, “The Stretched horizon and black hole com- plementarity,” Phys. Rev. D  48, 3743 (1993) [hep-th/9306069].
C. R. Stephens, G. ’t Hooft and B. F. Whiting, “Black hole evaporation without infor- mation loss,” Class. Quant. Grav. 11, 621 (1994) [gr-qc/9310006].


 \bibitem{FireWalls}
     A.~Almheiri, D.~Marolf, J.~Polchinski and J.~Sully,
   ``Black Holes: Complementarity or Firewalls?,''
  JHEP {\bf 1302}, 062 (2013).

\bibitem{Ashtekar1}
  A.~Ashtekar, V.~Taveras and M.~Varadarajan,
  ``Information is Not Lost in the Evaporation of 2-dimensional Black Holes,''
  Phys.\ Rev.\ Lett.\  {\bf 100}, 211302 (2008)
  [arXiv:0801.1811 [gr-qc]].

\bibitem{Bojowald} M.~Bojowald,
  ``Information loss, made worse by quantum gravity,'' [arXiv:1409.3157 [gr-qc]].

 \bibitem{wormhole} 
  J.~Maldacena and L.~Susskind,
   ``Cool horizons for entangled black holes,''
  Fortsch.\ Phys.\  {\bf 61}, 781 (2013)
  [arXiv:1306.0533 [hep-th]]. 
			
		\bibitem{planck-star} 
  C.~Rovelli and F.~Vidotto,
   ``Planck stars,''
  [arXiv:1401.6562 [gr-qc]].
	
	\bibitem{Mathur2} S. D. Mathur, ``How fuzzballs resolve the information paradox," Journal of Physics: Conference Series 462 
	(2013) 012034.
	
\bibitem{Okon1} ``Benefits of Objective Collapse Models for Cosmology and
Quantum Gravity" E.  Okon  \&  D. Sudarsky, {\it Found. of Phys.} {\bf 44}  114-143,  (2014);
 arXiv:1309.1730v1 [gr-qc] .


\bibitem{QFTinCS}
L. Parker and D. Toms, ``Quantum Field Theory in Curved Spacetime,'' Cambridge University Press (2007);
R. M. Wald, ``Quantum Field Theory in Curved Spacetime and Black Hole Thermodynamics,'' The University of Chicago Press (1994).

\bibitem{Page} D. N. Page and C. D. Geilker, {Phys. Rev. Lett.} {\bf 47},  979 (1981).	

\bibitem{Carlip} Is Quantum Gravity Necessary?
S. Carlip,  {Class. Quant. Grav.} {\bf 25}  154010 (2008).

   \bibitem{Collapse-and-Inflation} A. Perez, H. Sahlmman and D. Sudarsky, ``On the Quantum Mechanical Origin of the Seeds of Cosmic Structure," Classical and Quantum Gravity  {\bf 23}, 2317 (2006); D. Sudarsky, ``Shortcomings in the Understanding of Why Cosmological Perturbations Look Classical,"
International Journal of Modern Physics D {\bf 20}, 509
(2011) [arXiv:0906.0315 [gr-qc]]; S. J. Landau, C. G. Scoccola and D. Sudarsky, ``Cosmological
constraints on nonstandard inflationary quantum collapse models,"
  Physics Review D {\bf 85}, 123001 (2012)
 [arXiv:1112.1830 [astro-ph.CO]]; G. Le\'on Garc\'{\i}a, S. J. Landau and D. Sudarsky, ``Quantum Origin of the Primordial Fluctuation Spectrum and its Statistics,"
Physics Review D {\bf 88},  023526 (2013) [arXiv:1107.3054 [astro-ph.CO]];
 P. Ca\~nate, P. Pearle and D. Sudarsky,
``CSL Quantum Origin of the Primordial Fluctuation,"
Physics Review D {\bf 87}, 104024 (2013) [arXiv:1211.3463[gr-qc]].



\bibitem{israel} W. Israel, Nuovo Cim. 44B, 1 (errata in 48B, 463) (1966); 



  \bibitem{Alberto} A. Diez-Tejedor and D. Sudarsky, ``Towards  a formal  description  of  the collapse  approach to  the inflationary origin of  the seeds  of cosmic  structure,"
JCAP {\bf  045}, 1207 (2012)
[arXiv:1108.4928  [gr-qc]]. 


\bibitem{Sujoy 1} 
  S.~K.~Modak, L.~Ort\'iz, I.~Pe\~na and D.~Sudarsky,
  ``Non-Paradoxical Loss of Information in Black Hole Evaporation in a Quantum Collapse Model,''
  Phys.\ Rev.\ D {\bf 91}, no. 12, 124009 (2015)
  [arXiv:1408.3062 [gr-qc]].
	
	
	\bibitem{Sujoy 2} 
  S.~K.~Modak, L.~Ort\'iz, I.~Pe\~na and D.~Sudarsky,
  ``Black hole evaporation: information loss but no paradox,''
  Gen.\ Rel.\ Grav.\  {\bf 47}, no. 10, 120 (2015)
   [arXiv:1406.4898 [gr-qc]].

 
\bibitem{Wald-Unruh} W. G. Unruh and R. M. Wald, ``On evolution laws taking pure states to
mixed states in quantum field theory,"  {\it  Phys.Rev.} {\bf  D 52}, 2176-2182 (1995).


   
 \bibitem{Peskin}
T. Banks, L. Susskind , M. E. Peskin, ``Difficulties for the Evolution of Pure States Into Mixed States, "  
 {\it  Nucl.Phys. B} {\bf 244}  125 (1984). 



 \bibitem{Pearle:76} P. Pearle, ``Reduction of the state vector by a nonlinear Schr\"odinger equation," Phys. Rev. D {\bf 13}, 857
(1976).


  \bibitem{Pearle:79} P. Pearle, ``Towards explaining why events occur," Int. J. Theor. Phys. {\bf 18}, 489 (1979).


  \bibitem{GRW:85} G. Ghirardi, A. Rimini, T. Weber, ``A model for a unified quantum description of macroscopic and
microscopic systems," in A. L. Accardi (ed.) Quantum Probability and Applications, p. p. 223-232, Springer, Heidelberg (1985).


  \bibitem{GRW:86} G. Ghirardi, A. Rimini, T. Weber, ``Unified dynamics for microscopic and macroscopic systems,"
  Phys.Rev. D {\bf 34}, 470 (1986).

    \bibitem{Pearle:89} P. Pearle, ``Combining stochastic dynamical state-vector reduction with spontaneous localization,"
    Phys. Rev. A {\bf 39}, 2277-2289 (1989).

      \bibitem{GRW:90} G. Ghirardi, P. Pearle, A. Rimini, ``Markov-processes in Hilbert-space and continuous spontaneous
localization of systems of identical particles," Phys. Rev. A {\bf 42}, 7889 (1990).

 \bibitem{moreCSL} P. Pearle, ``Collapse models,"  [arXiv: quant-ph/9901077].





    \bibitem{measurement}
  D. Albert, Quantum Mechanics and Experience (Harvard University Press, 1992),
Chapters 4 and 5;
J. Bell, ``Quantum mechanics for cosmologists",  in Quantum Gravity II, Oxford
University Press, 1981;
D. Home, Conceptual Foundations of Quantum Physics: an overview from modern
perspectives (Plenum, 1997). Chapter 2;
  E. Wigner, ``The problem  of Measurement,'' Am. J. of Physics
{\bf 31}, 6 (1963);  A. Lagget, ``Macroscopic quantum Systems and the quantum theory of measurement,''
Prog.  Theor. Phys. Suppl. {\bf 69}, 80 (1980); J. S. Bell, ``Speakable and Unspeakable in Quantum Mechanics,"
(Cambridge University Press 1987); J. S. Bell, ``Against 'Measurement," Phys. World {\bf  3}, 33 (1990).

   \bibitem{More-measurement}    For reviews about the various approaches to
the measurement problem in quantum mechanics see, for instance,
 the classical reference M. Jammer, ``Philosophy of quantum mechanics.
 The interpretations of quantum mechanics in historical perspective,''	
(John Wiley and Sons, New York  1974);  A. Peres, ``Quantum Theory: Concepts and Methods" (Kluwer, Academic Publishers, 1993); R. Omnes, ``The Interpretation of Quantum Mechanics," (Princeton University Press 1994); and the more specific critiques
S. L. Adler `` Why Decoherence has not Solved the Measurement Problem: A Response to PW Anderson," Stud. Hist. Philos. Mod. Phys. {\bf 34}, 135-142 (2003) [arXiv: quant-ph/0112095]; A. Elby, Why modal interpretations of quantum mechanics don't solve the measurement problem, Found. of Phys. Lett.
{\bf  6},  5-19 (1993).



\bibitem{Mau:95}T. Maudlin, ``Three measurement problems," Topoi {\bf 14}(1), 715 (1995).


 \bibitem{Interpretaciones}
  D. Durr, S. Goldstein, and N. Zangh, "Bohmian Mechanics and the Meaning of the
Wave Function," in Cohen, R. S., Horne, M., and Stachel, J., eds., Experimental
Metaphysics -- Quantum Mechanical Studies for Abner Shimony, Volume One;
Boston Studies in the Philosophy of Science 193, ( Kluwer Academic Publishers,
1997);
J. S. Bell, ``On the impossible pilot wave", Foundations of Physics 12 (1982), pp.
989-99;
D. Wallace, The Emergent Multiverse. Oxford University Press, 2012;
C. Fuchs and A. Peres, ``Quantum Theory Needs No ``Interpretation"". Physics Today
53(3) (2000), pp. 70-71;
O. Lombardi and D. Dieks, ``Modal interpretations of quantum mechanics", The
Stanford Encyclopedia of Philosophy, 2014;
E. Joos et al, Decoherence and the Appearance of a Classical World in Quantum
Theory, 2nd edition (Springer, 2003); W. Zurek, ``Decoherence and the transition from quantum to classical," Phys.
Tod., vol. 44, no. 10, 1991.


\bibitem{critiques}
A. Kent, ``Against Many-Worlds Interpretations", online at http://xxx.arxiv.org/abs/gr-qc/9703089;
H. Brown, and D. Wallace, ``Solving the measurement problem: de Broglie-Bohm
loses out to Everett". Foundations of Physics 35 (2005), pp.517-540;
J. Bub, Interpreting the Quantum World (Cambridge, 1997), chapter 8, pp. 212-236.
(Rather critical discussion of the decoherence-based approaches).

 
\bibitem{collapse model review}
A.~Bassi \& G.C.~Ghirardi, Phys.~Rept. {\bf 379}, 257 (2003).
A.~Bassi, K.~Lochan, S.~Satin, T.~P.~Singh, \& H.~Ulbricht, Rev.~Mod.~Phys.~{\bf 85}, 471 (2013)


\bibitem{BED1} D.~J.~Bedingham, ``Relativistic State Reduction Dynamics,'' Found.~Phys.~{\bf 41} 686 (2011). 

\bibitem{BED2} D.~J.~Bedingham, ``Relativistic state reduction model,
" J.~Phys.: Conf.~Ser.~{\bf 306} 012034 (2011).


\bibitem{relativistic collapse  models}
 R. Tumulka, ``A relativistic version of the Ghirardi-Rimini-Weber model," J. Stat. Phys. {\bf 125}, 821 (2006);
	P. Pearle 
"A Relativistic Dynamical Collapse Model"
 [arXiv:1412.6723 [quant-ph]].


  \bibitem{Penrose}  R. Penrose, ``The Emperor's New Mind,'' (Oxford
  University Press 1989); R. Penrose, ``On Gravity's  Role in Quantum State Reduction,''
  in Physics meets philosophy at the Planck scale, Callender, C. (ed.) (2001).
	
	\bibitem{penrose-new1} R. Penrose, ``On Gravity's Role in Quantum State Reduction,'' General Relativ. and Gravit., {\bf~8}, 581 (1996).
	
	  
  \bibitem{Penrose-BH-Collpase} R. Penrose, ``Time asymmetry and quantum gravity," in C. J. Isham, R. Penrose, D. W. Sciama, (eds.) Quantum Gravity II, p. 244 (1981).
	
	\bibitem{penrose-new2} R. Penrose, ``On the Gravitization of Quantum Mechanics 1: Quantum State Reduction,'' Found. Phys. {\bf 44}, 557-575 (2014); R. Penrose, ``On the Gravitization of Quantum Mechanics 2: Conformal Cyclic Cosmology,'' Found.~Phys., {\bf 44}, 873-890 (2014).
  
\bibitem{CGHS92}
      C. G. Callan, S. B. Giddings, J. A. Harvey, and A. Strominger,
      ``Evanescent black holes,"
      Phys. Rev. D \textbf{45}, R1005 (1992).
\bibitem{aFabjNs05}
      A. Fabbri and J. Navarro-Salas,
      Modeling Black Hole Evaporation
      (Imperial College Press, London 2005).


 \bibitem{cghs-more}      S. B. Giddings and W. M. Nelson, ``Quantum emission from two-dimensional black holes,''      Phys. Rev. D \textbf{46}, 2486 (1992).
 
\bibitem{dEspagnat}
B.~d'Espagnat. \emph{Conceptual Foundations of Quantum Mechanics}. (Addison-Wesley, 2nd. ed., 1976)

\bibitem{relativistic collapse  ideas 2}
G.C.~Ghirardi, R.~Grassi, P.~Pearle, Found.~Phys.~{\bf 20}, 1271 (1990)

       
\bibitem{BF} N.~D.~Birrell \& L.~H.~Ford, Ann.~Phys.~{122} 1 (1979).

\bibitem{BPP} T.~S.~Bunch, P.~Panangaden, \& L.~Parker, J.~Phys. A: Math.~Gen. {\bf 13} 901 (1980).


\bibitem{relativistic collapse  ideas 1}
W.~Myrvold, Stud.~Hist.~Phil.~Mod.~Phys.~{\bf 33} (2002), 435.


\bibitem{CAUSALSET}
L.~Bombelli, J.~Lee, D.~Meyer, \& R.~Sorkin, Phys.~Rev.~Lett.~{\bf 59}, 52 (1987).


\bibitem{GERMANIUM}
W.~Feldmann \& R.~Tumulka, J.~Phys.~A: Math.~Theor.~{\bf 45}, 06504 (2012).


  
  
\bibitem{Genesis} 	
E.  Okon \&  D. Sudarsky, ``A (not so?) novel explanation for the very special initial state of the universe,"   arXiv:1602.07006 [gr-qc] .

 \bibitem{LIV}

 J. Collins, A. Perez, D. Sudarsky, L. Urrutia, and
H. Vucetich, ``Lorentz Invariance in Quantum Gravity: A New fine tunning problem?," {\it Phys. Rev. Lett.} {\bf 93}, 191301, (2004);  C. Rovelli and S. Speziale, ``Reconcile Planck-scale discreteness and the Lorentz-Fitzgerald    contraction, {\it Phys. Rev.} {\bf D 67}, 064019 (2003) [arXiv:gr-qc/0205108];  F. Dowker and R. Sorkin, ``Quantum Gravity Phenomenology, Lorentz Invariance and Discreteness,''
   [arXiv:gr-qc/0311055].
  
\bibitem{RST-Us} S. K. Modak, L. Ort\'iz,  I. Pe\~na \&  D. Sudarsky,  In preparation.

\bibitem{Bonilla-Senovilla}  M. Bonilla \&  J. Senovila, Gen. Rel.   \& Grav.   {\bf  29}, 91  ( 1997).

	
 
    



 


   








\end{thebibliography}
\end{document}